\documentclass[iop,twocolappendix]{emulateapj}
\usepackage[utf8]{inputenc}
\usepackage[margin=1.0in]{geometry}
\usepackage[utf8]{inputenc}
\usepackage{graphicx}
\usepackage{amssymb}
\usepackage{epstopdf}
\usepackage{wasysym}
\usepackage{natbib}
\usepackage{textcomp}
\usepackage{verbatim}
\usepackage{hyperref}
\begin{document}

\title{Empirically Constrained Predictions for Metal-Line Emission from the Circumgalactic Medium}
\author{Lauren Corlies, David Schiminovich}
\affil{Department of Astronomy, Columbia University, New York, NY 10027, USA}

\begin{abstract}
The circumgalactic medium (CGM) remains one of the least constrained components of galaxies and as such has significant potential for advancing galaxy formation theories.  In this work, we vary the extragalactic ultraviolet background for a high-resolution cosmological simulation of a Milky Way-like galaxy and examine the effect on the absorption and emission properties of metals in the CGM.  We find that a reduced quasar background brings the column density predictions into better agreement with recent data.  Similarly,  when the observationally derived physical properties of the gas are compared to the simulation, we find that the simulation gas is always at temperatures approximately 0.5 dex higher. Thus, similar column densities can be produced from fundamentally different gas.  However, emission maps can provide complementary information to the line-of-sight column densities to better derive gas properties. From the simulations, we find that the brightest emission is less sensitive to the extragalactic background and that it closely follows the fundamental filamentary structure of the halo.  This becomes increasingly true as the galaxy evolves from $z=1$ to $z=0$ and the majority of the gas transitions to a hotter, more diffuse phase.  For the brightest ions (C{\scriptsize III}, C{\scriptsize IV}, O{\scriptsize VI}), detectable emission can extend as far as 120 kpc at $z=0$.   Finally, resolution is a limiting factor for the conclusions we can draw from emission observations but with moderate resolution and reasonable detection limits, upcoming instrumentation should place constraints on the physical properties of the CGM.
\end{abstract}

\maketitle

\section{Introduction}
Perhaps the most basic process of galaxy formation, the flow of gas into and out of a galaxy, remains as one of the least understood.  The key seems to lie in our lack of understanding of the circumgalactic medium (CGM). Roughly defined as the gas surrounding galaxies at 10 to 300 kpc, the CGM encompasses all gas in transition: gas falling onto the galaxy for the first time; gas that is being driven out by multiple feedback processes; gas that is being stripped from infalling satellite galaxies; and gas that is currently being recycled by the galaxy \citep[see][for review]{Putman_review}.

The structure of this gas halo depends on the mass and redshift of the galaxy in question. Currently, gas is thought to be accreted through two main modes - a ``hot'' mode where the gas is shock heated as it enters the halo and a ``cold'' mode where the gas remains in unshocked filamentary structures that can potentially penetrate all the way to the disk \citep{Keres_2005,Fumagalli_2011}.  Milky Way-like galaxies are thought to transition from the cold mode to the hot mode by the present day but the details of this transition are neither theoretically agreed upon nor well-constrained observationally \citep{Brooks_2009,ryan,Nelson_2015b}.

In addition to these inflows, the outflow of gas from the galaxy is equally important in shaping the CGM \citep{Nelson_2015a, Marasco_2015,Suresh_2015}. Stellar feedback of some form is clearly needed to prevent the overcooling of gas and the formation of unrealistic stellar bulges in simulations \citep{Agertz_2011,Brook_2011,Hummels_2012}. It is also the most effective way of enriching the IGM to the non-pristine levels that are observed \citep{Oppenheimer_2008, Wiersma_2010,Barai_2013,Ford_2013}.  While such outflows are regularly seen, the exact physical process driving them and the extent of their influence is uncertain \citep{Turner_2015}.  Multiple preferred forms of Type II supernovae feedback are implemented and recent work has begun to implement more detailed processes such as radiation pressure from supernovae \citep{Hopkins_2012,Agertz_2013,Ceverino_2014,Trujillo_Gomez_2015}, cosmic rays \citep{Booth_2013,Salem_2014a,Salem_2014b}, AGN \citep{Sijacki_2007,Booth_2009}, and direct modeling of a kinetic energy component \citep{Simpson_2015} to name a few.  In short, putting constraints on these many models is fundamental to furthering our understanding of galaxy formation.  

In general, cosmological galaxy simulations are tuned to reproduce global and primarily \emph{stellar} properties of galaxies such as the stellar mass function and the star formation rate density function \citep{Dave_2011,schaye_eagle, Nelson_2015a}. Another benchmark is the creation of thin, extended stellar disks \citep{Governato_2007}.  The H{\scriptsize I} mass function is a constraining gas property but again looks at the total mass and not its distribution throughout the galaxy. \citep{Dave_2013}  Recently, theoretical work has begun to compare the simulated CGM to column densities and equivalent width measurements as a function of impact parameter from the center of the galaxy \citep{hummels, ford, Liang_2015, oppenheimer_2016}. The majority of the simulations have difficulty in matching the large amount and high covering fraction of O{\scriptsize VI} measurements, tracing the hottest gas phase (except recently for high-mass galaxies \citep{Suresh_2015b} and with cosmic ray feedback \citep{Salem_2015}).  Their success varies when looking at cooler, less ionized lines (Mg{\scriptsize II}, C{\scriptsize III}, Si{\scriptsize IV} etc.) but in general, the data reveal large amounts of metal-enriched gas at large impact parameters that is hard to reproduce theoretically.  In this way, measurements of the CGM can put strong restrictions on feedback models, independent of the global properties that are already used.

The most successful method of observing the CGM is in the absorption lines of quasar spectra.  At higher redshifts, Lyman $\alpha$ and the ultraviolet metal lines of interest have shifted into the optical, making observations easier and successful \citep{Steidel_2010,Simcoe_2004}. At low-redshift, several studies have begun pushing our knowledge of the more local CGM with measurements of Mg{\scriptsize II} \citep{Chen_2010} and O{\scriptsize VI} for a number of galaxies \citep{Prochaska_2011, Thom_2008}.  The recent installation of the Cosmic Origins Spectrograph (COS) on HST has enabled a new survey of the CGM of low-redshift ($z \approx 0.2$), massive, isolated galaxies. The COS Halos Survey has provided a large, uniformly measured sample of the H{\scriptsize I} column densities \citep{Tumlinson_HI}, metal line absorption \citep{werk13}, and O{\scriptsize VI} column densities \citep{tumlinson_OVI}. As accretion and outflows are expected to vary with redshift in addition to mass, low-redshift studies such as these are crucial as is the need to push to even lower redshifts.

A complementary approach is to observe the CGM $\emph{directly}$ in emission.  Quasar spectra will always be limited by the small number of sight lines through each galaxy.  An emission map has the potential to provide insight into the physical state of an entire galaxy halo.  While promising, the low density of the gas has made this observation challenging.  The most success has come from high-redshift surveys for Lyman $\alpha$ emitters \citep[e.g.][]{Bridge_2013,Gawiser_2007} and the more extended Lyman $\alpha$ blobs/halos \citep[e.g.][]{Matsuda_2011, Steidel_2011, Steidel_2000} but metal-line emission has remained elusive \citep{Battaia_2015}.  Recently, the development of new integral field units, MUSE and CWI (and its successor KCWI), now allows for a study of the kinematics of the gas. Early work has already suggested that the absorbers can be linked to global outflows \citep{Swinbank_2015} as well as filamentary inflows \citep{Martin_2014}.  At low redshift, the upcoming FIREBall-2 is building upon its predecessor \citep{Milliard_2010} and pushing the boundaries of low surface brightness UV observations. This, in addition to any small or large near-future UV space telescope mean that direct UV observations of the CGM are closer than ever.

With these advancements in mind, this work looks to take advantage of new data while preparing for future observations. We take a high-resolution, cosmological, hydrodynamical simulation of a Milky Way-like galaxy and compare it to recent column density data.  We then ask what emission we could presume to detect with upcoming facilities.

Previous studies of this same simulation provide a solid foundation for this work. \citet{ximena} demonstrated that in-falling satellites provide much of the cold, high metallicity gas found in the halo at $z=0$ whereas \citet{ryan} quantified how much gas of a given temperature is accreted at low-$z$. This existing physical insight allows us to better understand the evolution of the CGM and the contribution of different accretion modes. 

In this paper, we look to build on this work when interpreting our emission predictions.  In Section 2, basics of the simulation used and the photoionization model are summarized.  In Section 3, the simulation is compared to column density observations to put empirical constraints on the interpretation of the simulation. In Section 4, the emission signatures of this gas and how they evolve are examined and its observational properties are explored. Finally, the broader context of the work is discussed in Section 5 and the results are summarized in Section 6.

\section{Methodology}
\subsection{Simulation Basics}
We analyze the cosmological, hydrodynamical simulation of \citet{ryan} performed with {\tt enzo}, an Eulerian, adaptive mesh refinement, hydrodynamical code \citep{enzo}. A Milky Way-like halo was identified from within an initial low-resolution run with a periodic box of $L = 25\ h^{-1}$ Mpc comoving on a side with cosmological parameters consistent with WMAP5. This galaxy was centered in a box of length $\approx 5\ h^{-1}$ Mpc which was then resimulated with 10 levels of refinement.  The selected galaxy has a final halo mass of $1.4 \times 10^{12} M_{\astrosun}$ and contains over 8.2 million dark matter particles within its virial radius, with $m_{\mathrm{DM}} = 1.7 \times 10^5 M_{\astrosun}$. The final stellar mass is $1.9 \times 10^{11} M_{\astrosun}$, placing the halo above the M$_{\mathrm(star)}$-M$_{\mathrm(halo)}$ relation as is common with simulations of this type \citep{guo_2010}.  The maximum spatial resolution stays at 136-272 pc comoving or better at all times.

The simulation includes metallicity-dependent cooling, a metagalactic UV background, shielding of UV radiation by neutral hydrogen, and a diffuse form of photoelectric heating. The code simultaneously solves a complex chemical network involving multiple species (e.g. H{\scriptsize I}, H{\scriptsize II}, H$_2$, He{\scriptsize I}, He{\scriptsize II}, He{\scriptsize III}, $e^-$) and metal densities explicitly.  

Star formation and stellar feedback are included in the simulation. Star particles have a minimum initial mass of $m_* = 1.0 \times 10^5 M_{\astrosun}$ and are created if $\rho > \rho_{\mathrm{SF}}$ and with a violation of the Truelove criterion. Supernovae feedback is modeled following \citet{Cen_2005}, with the fraction of the stellar rest-mass energy returned to the gas as thermal energy, $e_{\mathrm{SN}} = 10^{-5}$, consistent with the \citet{Chabrier_2003} initial mass function.  The metal yield from stars, assumed to be 0.025, represents metal production from supernovae of both Type Ia and Type II. This metallicity is traced as a single field and abundances are generated throughout the paper assuming the solar abundance. Feedback energy and ejected metals are distributed into 27 local cells centered at the star particle in question, weighted by the specific volume of the cell. The metals and thermal energy are released gradually, following the form: 
$f(t,t_i,t_*) = (1/t_*)[(t - t_i)/t_*] \exp[-(t - t_i)/t_*]$, where $t_i$ is the formation time of a given star particle, and $t_* = \mathrm{max}(t_{\mathrm{dyn}}, \ 3 \times 10^6 \mathrm{yr})$ where $t_{\mathrm{dyn}} = \sqrt{3\pi / (32 G \rho_{\mathrm{tot}})}$ is the dynamical time of the gas from which the star particle formed. The metal enrichment inside galaxies and in the IGM is followed self-consistently in a spatially resolved fashion.  For details of these prescriptions, we direct the reader to \citet{ryan}.

\subsection{Ionization Modeling}
To calculate the relevant ionization processes of interest, the simulation was post-processed with the photoionization code \textsc{cloudy} \citep[version 10.0, last described in][]{cloudy} in conjunction with the cooling map generation code \textsc{roco} \citep{smith_2008} and the simulation analysis suite {\tt yt} \citep{yt}. For each model discussed in the upcoming sections, the following procedure was used to produce the column density and emission predictions.

First, \textsc{cloudy} look-up tables of ion fractions and emissivity were constructed for a given ionization background as a function of temperature ($10^3 < T < 10^8, \ \Delta \log_{10} T = 0.1$) and hydrogen number density ($10^{-6} < n_{\mathrm{H}} < 10^2,\  \Delta \log_{10} n_{\mathrm{H}} = 0.5$). Each table assumes solar metallicity and abundances.   The grid is then interpolated for every cell to the correct temperature and $n_{\mathrm{H}}$.  Then, $n_{Xi}$, the number density of given ionization state of element $X$ (C{\scriptsize III}, Si{\scriptsize IV}, O{\scriptsize VI}, etc.) is calculated as:
\begin{equation}
n_{Xi} = n_{\mathrm{H}}(n_X/n_{\mathrm{H}})(n_{Xi}/n_X)
\end{equation}
where $(n_X/n_{\mathrm{H}})$ is the elemental abundance relative to hydrogen and $(n_{Xi}/n_X)$ is the ion fraction computed by \textsc{cloudy}.  Here, the elemental abundance is given as the solar abundance scaled by the metallicity reported in the simulation. 
The emissivity is more straightforward as \textsc{cloudy} directly reports the emissivity at a given temperature and density that is then again scaled by the metallicity.

With these number densities and emissivities, producing the corresponding column density and surface brightness values is done as projections through the simulation with {\tt yt}.  Throughout the paper, we assume a box that is 320 kpc across and 500 kpc deep, ensuring the selection of gas associated with the galaxy.  Each projection and radial profile is made with a resolution of 1 physical kpc, unless otherwise stated.

Finally, throughout the paper, the assumed ionization field, the extragalactic ultraviolet background (EUVB), is varied to examine the agreement of the simulation predictions with the column density measurements. To this end, we take the 2005 updated version of the \citet{HM01} background of \textsc{cloudy} and split it into its two components - quasars and galaxies. Then, the intensity of each component can be varied and the changes in the predicted column densities studied.  The quasar component dominates at short wavelengths and is responsible for the majority of the ionizing radiation in the calculations. In this way, varying the quasar component has more significant consequences than varying the galactic component.   A more detailed discussion of the differences among these backgrounds is found in the Appendix.

Because this is a post-processing of the simulation, this technique is not fully self-consistent. It does not capture the underlying effects on the temperature and density that arise from changing the ionization background used in computing the cooling of the gas. However, the overall galaxy evolution and supernova feedback are thought to dominate the evolution of the gas density, temperature and metallicity more than the choice of EUVB and the ion fractions of interest here are less important in determining these large-scale properties. These limitations remain as part of the uncertainty in the following calculations but the overall conclusions should be robust.  

However, the response of the simulation to the changes in the EUVB reflects the field's true influence on the ionization state of the simulated gas, assuming ionization equilibrium. The field is a fundamental property of the physics of the calculation used in calculating the ion fractions and emissivity in \textsc{cloudy}. The density and temperature of the gas are not expected to vary much with the choice of EUVB, as discussed in the Appendix.

\section{Absorption}
In this section, we look to place the simulation in the context of a set of current absorption-line observations.  First, column density maps of a series of ions are generated and the resulting CGM structure is analyzed. Next, the reliability of the simulation is tested by examining its agreement with available observations, specifically the COS Halos survey \citep{werk13}. \\

\begin{figure}[t]
\centering
\includegraphics[width=0.3\textwidth]{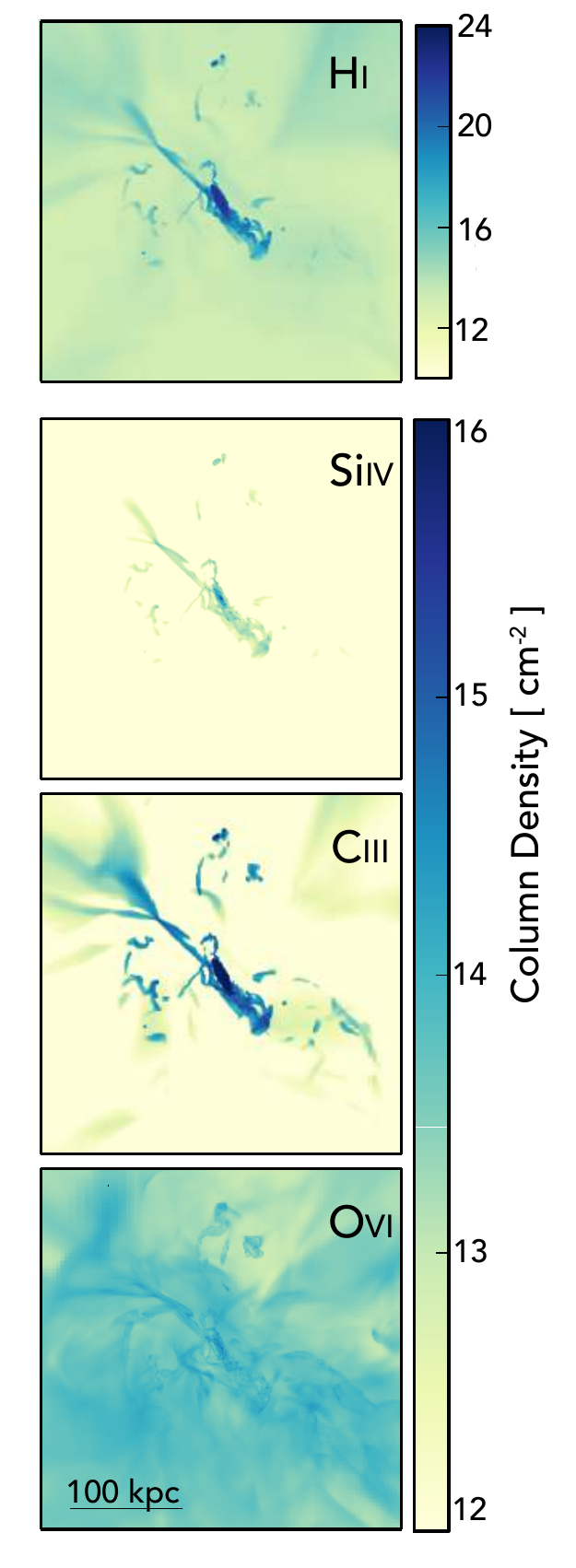}
\caption{Column density maps of H{\scriptsize I}, Si{\scriptsize IV}, C{\scriptsize III}, and O{\scriptsize VI} respectively at $z=0.2$ with a resolution of 1 kpc. Note H{\scriptsize I} has a unique color bar.  O{\scriptsize VI} has the largest covering factor with moderately high column densities extending uniformly over 100 kpc. C{\scriptsize III} has a smaller covering factor but reaches higher densities in visible filaments and stripped satellite material. Si{\scriptsize IV} is the weakest as its peak ionization temperature is slightly below the typical temperature of the halo gas.   \label{abs_proj.fig}}
\end{figure}

\subsection{Column Density Maps}

In order to better visualize the CGM column density distribution of the simulated galaxy, we first examine Figure \ref{abs_proj.fig}, which shows column density maps for four ions at $z=0.2$ with a resolution of 1 kpc.  This set of ions allows us to probe from the coldest gas (H{\scriptsize I}) to the hottest gas (O{\scriptsize VI}) and the warm gas in between (Si{\scriptsize IV}, C{\scriptsize III}).  What is first apparent is the intricate structures visible for all of these ions. The H{\scriptsize I} naturally has the largest column densities in the filaments that trace the high density structures within the gas.  

The distribution of Si{\scriptsize IV} and C{\scriptsize III} closely follows that of the H{\scriptsize I}.  The greater strength of C{\scriptsize III} is related to the fact that it is approximately 10 times more abundant than Si{\scriptsize IV} for a given metallicity.  These low ionization ions are found mostly in the higher density gas because the high average temperature of the CGM outside these regions prefers higher ionization states. These trends are also true of other low ions, such as Si{\scriptsize III}, which show similar features.  Conversely, in this map, although the O{\scriptsize VI} does retain traces of the same underlying structures seen as slightly enhanced column density regions, its higher ionization energy allows it to exist in hotter gas. In this way, the O{\scriptsize VI} has both the largest extent and obtains an appreciable column density value for almost the entire area of the map.  This is consistent with \citet{tumlinson_OVI} who found O{\scriptsize VI} in all of their star forming galaxies, implying a high covering fraction as seen here.

\subsection{Comparison to COS Halos Column Densities}
With column density maps in hand, we now compare the simulation to the uniform, galaxy-selected, quasar sample of COS Halos data, which provides measurements of the column densities of multiple ions as a function of impact parameter to low sensitivities.  The ions presented here (SiIII, SiIV, CIII, OVI) span a wide range in ionization energy while having a large number of observations in the COS Halos sample. CIV is excluded as observations are limited by the degraded sensitivity of COS for wavelengths $\lambda > 1500$ \AA, necessary for this redshift sample \citep{werk13}.  As we are considering a single galaxy, the simulation is not expected to reproduce every aspect of the larger population sampled by the survey.  Because of this fact, coupled with the large number of upper and lower limits in the data, the comparison made here between simulated and observed column densities is visual.  Every pixel in the column density map is shown so that the validity of the conclusions drawn here are easily confirmed.

\begin{figure*}
\centering
\includegraphics[width=0.7\textwidth]{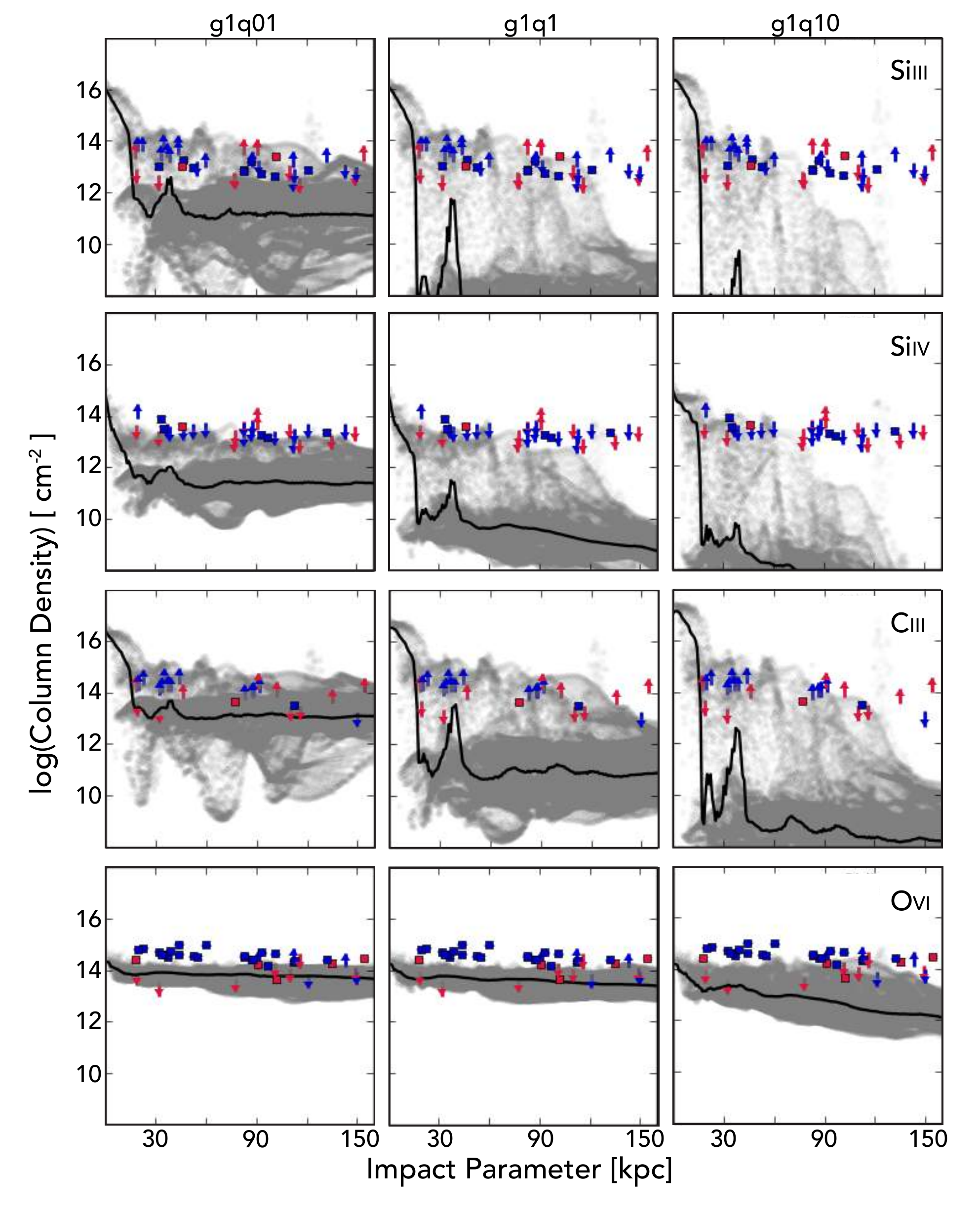}
\caption{Radial profiles of four different ions with gray points showing the values of individual pixels and the black line showing the median. Overplotted are column densities from the COS Halos survey, colored as either star-forming (blue) or passive (red). The fiducial UV background (HM05) fails to reproduce the observed column densities except for the peaks from the remaining filamentary structure and satellite galaxies.  Lowering the background to 0.01 times the normal quasar intensity provides a better match to Si{\scriptsize IV} and C{\scriptsize III} though some tension remains.  O{\scriptsize VI} appears more collisionally ionized such that the changes are less marked but no background successfully reproduces the high O{\scriptsize VI} absorption seen in star-forming galaxies.   \label{coldens_scatter.fig}}
\end{figure*}

As a base case, we assume the standard HM05 background, labeled as g1q1 in figures.  Using the method described in Section 2, the column densities are computed for each ion as a function of impact parameter from the galactic center at $z=0.2$, the approximate redshift of the data.  The center panes of Figure \ref{coldens_scatter.fig} show the resulting average radial profiles of three projection angles as well as the value of each pixel for a single projection (shown in Figure \ref{abs_proj.fig}). Each pixel has a width of 1 kpc. The data points are detections and upper and lower limits respectively of the COS Halos data set.  The color of the data point indicates whether the galaxy is considered star forming (with sSFR $> 10^{-11}$ yr$^{-1}$) or passive as in \citet{werk13}.  At $z=0.2$, the simulated galaxy has a stellar mass of $1.9 \times 10^{11} M_{\astrosun}$ and star formation rate (SFR) of $6.22 \ M_{\astrosun} / yr$, making it a star-forming galaxy by this classification as expected.  This stellar mass is typical of a COS Halos galaxy but this SFR leads to a rate at $z=0$ that is high compared to the actual Milky Way \citep[noted by][]{ximena, ryan}.  However, in the COS Halos sample, 5 galaxies have this SFR or higher. These points show no special trends in the column densities \citep{werk13} so the comparison done here is valid. 

These plots highlight both the average trends of the halo gas as well as the structures seen in the column density maps of Figure \ref{abs_proj.fig}.  In general, the median column densities remain roughly constant with impact parameter.  However, filamentary structures and satellite galaxies (the peak seen around 40 kpc) provide the possibility of a quasar sightline measuring higher than average column densities.  Furthermore, the distribution of the pixel values does vary somewhat with the projection angle.  In particular, a face-on projection reduces the scatter in the inner radii as the disk dominates the gas distribution.  However, the average values are generally unaffected.  Throughout the paper, we plot a mostly edge-on projection which allows for a better evaluation of how gas extends perpendicularly from the disk while lessening the influence of the disk itself.

It is apparent that for this simulation, this model is not a good fit to most of the data. The higher density filamentary structures bring many of the C{\scriptsize III} measurements into alignment with the data but the small covering fraction of these filaments makes it unlikely that they constitute a large fraction of the COS Halo absorber population.  The Si{\scriptsize IV} data is composed of many upper limits which means the low predicted values may be more in-line with the simulation. However, the detections are still mostly too high to match the simulation at the larger impact parameters. 

The O{\scriptsize VI} measurements, on the other hand, are comprised mostly of detections.  Our inability to match the O{\scriptsize VI} prediction for the star-forming galaxy model highlights a true disagreement.  The observed galaxies with similarly high star formation rates as our simulated galaxy have properties similar to the other star-forming galaxies while the simulation is more in agreement with the passive population.  Taken together, this suggests that the details of the hot phase of the CGM are not being properly reproduced.

Nevertheless, it is encouraging that the simulated gas shows a roughly flat radial profile like the data, as this was not guaranteed a priori. Figure \ref{radial_profiles.fig} shows the radial profiles of the density, temperature, and metallicity at z=0.2 in orange. The density decreases much faster than temperature beyond the disk (excepting peaks which represent satellite galaxies) while the metallicity actually begins to rise beyond 50 kpc. The combination of warm/hot temperatures, falling density, and increasing metallicity combine to produce the roughly flat column densities seen here in Figure \ref{coldens_scatter.fig}.  Yet, the points are not as tightly clustered as the data and are roughly two orders of magnitude too low.  

It is tempting to simply increase the metallicity of the gas to increase the column densities but it is not clear that this would lessen the discrepancies with the observations. The metallicity and temperature of the gas are intimately linked. Increasing the metallicity of the gas may lower the temperature such that the simulation remains in disagreement with some of the data.  For example, Si{\scriptsize III} and Si{\scriptsize IV} which prefer colder temperatures, would most likely benefit from added metallicity but O{\scriptsize VI} which prefers hotter temperatures may not. Furthermore, keeping the density and temperature fixed, the metallicity would need to be raised by approximately two orders of magnitude to bring better agreement with the data, assuming the fiducial HM05 EUVB.  This would put much of the CGM at solar metallicity or above, in contrast to most expectations and the measurements of \citet{werk14}. 

Instead, these discrepancies are an indication that the simulation which is tuned to reproduce bulk stellar properties of galaxies over time fails to do the same for these multi-phase CGM gas properties. However, this is not the first simulation to have such issues.  \citet{hummels} also analyze an {\tt enzo} simulation with a similar thermal feedback prescription but at lower resolution and report the same difficulties.  Likewise, the SPH simulation of \citet{ford} also fails to reproduce the O{\scriptsize VI} densities even though they implement a non-thermal wind prescription for their feedback.  One success is that of \citet{Salem_2015}, whose implementation of cosmic ray feedback successfully matches the data for all ions.  A discussion of these different methods is found in Section 5.

Thus, the typical solution that is invoked and explored in these works and many others is a modification to the stellar feedback prescription, changing the density, temperature and metallicity of the simulated CGM \citep[e.g. OWLS, EAGLE, FIRE described in][respectively]{schaye_owlsim, schaye_eagle,hopkins_fire}. This range of parameterizations can have an uncertain impact on the gas quantities such that new feedback solutions require the simulation to be re-run to capture the changes.  We explore the role of feedback further in the Discussion Section.

However, the EUVB is also important in setting the ionization state of the gas and is not well constrained. Variations of the \citet{HM96} background (e.g. HM96, HM01, HM05, HM12) are implemented in most simulations and \textsc{cloudy}. While the best models to date, there is still significant uncertainty in the exact strength and shape of the EUVB.  To this end, we chose to explore the impact of this uncertainty by varying the intensity of the galaxy and quasar components of the HM05 background and examining the effects on the resulting column densities.  In the following analysis, two bracketing cases of the quasar intensity are presented to examine the reasonable range of effects on the predicted column densities.  At one end, the quasar intensity is 100 times less intense than standard (g1q01) and at the other, the quasar intensity is ten times more intense (g1q10).  These properties are summarized in Table \ref{gq.tab} for easy reference.  We performed the same analysis for a range of quasar intensities spanning these two cases and the trends seen across the three values presented here are consistent with these results.  Furthermore, these two cases bracket current estimates of the photoionization rate with high redshift Lyman $\alpha$ forest studies preferring higher backgrounds \citep{kollmeier_underproduction,shull_2015} and with low-redshift H$\alpha$ upper limits preferring lower backgrounds \citep{adams_2011}.  Further discussion of the EUVB and its uncertainty can be found in the Appendix.  In addition, the galaxy intensity was varied in a similar way but with little to no effect on the predicted column densities as it provides less of the ionizing flux.  

\begin{table}[h]
\caption{EUVB Model Summary \label{gq.tab}}
\begin{center}
\begin{tabular}{ c|c|c } 
 \hline
Name & Galaxy & Quasar \\ \hline 
 g1q01 & 1.0 & 0.01 \\ 
 g1q1 & 1.0 & 1.0 \\ 
 g1q10 & 1.0 & 10.0 \\ 
 \hline
\end{tabular}
\end{center}
\end{table}

The first and last columns of Figure \ref{coldens_scatter.fig} once again show the radial profiles of the column densities of our simulated galaxy but with these altered EUVBs.  With g1q01 in the first column, it appears that lowering the quasar intensity to 0.01 times its normal value provides a much better fit to the low-ion data, Si{\scriptsize III}, Si{\scriptsize IV}, and C{\scriptsize III}. The majority of the pixels are now in better agreement with the data which is consistent with the idea that this softer spectrum is no longer over-ionizing the gas.  Raising the quasar intensity as seen in the last column with the g1q10 models results in a much larger disagreement between the simulation and the data, which is thus consistent with the picture of over-ionization. Together, this suggests that photoionization is the dominant process in producing these low ions.  O{\scriptsize VI}, on the other hand, is mostly unaffected, suggesting that the gas is predominantly collisionally ionized.  Most of the halo volume is at about the same density and temperature, accounting for the small spread in O{\scriptsize VI} column density values.  The scatter that is introduced is actually towards lower column densities with larger quasar intensity, consistent with the recombination rate of the lower density gas not being able to counterbalance the increased photoionization.  This demonstrates that producing the correct amount of O{\scriptsize VI} is not a simple matter of increasing the photoionization of the CGM. 

Observationally, there is support for the approach of varying the EUVB. \citet{crighton} allowed the power law slope of the \citet{HM12} (HM12) background between 1-10 Rydbergs to vary and found that half the components in their absorption spectra preferred an altered slope. One component agrees with the findings here, preferring a softer background, but the others are better fit by a slightly harder spectrum.  If variations of the EUVB are necessary to explain absorption components within the same sightline, it is reasonable to expect that the EUVB would vary amongst the many galaxies that comprise the COS Halos sample. Examining how this variation changes simulated predictions given otherwise identical physical conditions can thus provide insight into how to interpret such measurements.

On the other hand, the preference of this simulation for a weaker EUVB background is in fact in contrast both with these column density component measurements as well as the known limitation of the HM12 background failing to reproduce the column density distribution of Lyman $\alpha$ forest absorbers, known as the photon underproduction crisis \citep{kollmeier_underproduction,shull_2015}.  Solving this crisis calls for an increase in the photoionization rate of the HM12 background. However, the HM05 model used here is more consistent with the findings of \citet{kollmeier_underproduction} while it is less consistent with the log(N$_{\mathrm{H{\scriptsize I}}}$) $> 14.0$  distribution plotted in \citet{shull_2015}.  This uncertainty in the low-redshift EUVB supports our decision to vary its intensity though in light of the on-going efforts in feedback and sub-grid physics, we acknowledge that this is likely not enough to bring full agreement between the simulation and observations.  If a different feedback method allowed for the gas in the simulation to be cooler at late times, the EUVB would not have to be so low to produce the necessary amount of low ions such as Si{\scriptsize IV} and C{\scriptsize III}. 

Overall, we find that column densities predicted by the simulation using the standard HM05 background do not provide the best match to recent observations. Instead, the simulation prefers a softer, reduced quasar intensity to produce the necessary large amount of low ionization gas.  This demonstrates that simulation predictions are sensitive to the assumed EUVB and that this assumption should be considered in conjunction with efforts to vary feedback methods but that this preference of a reduced EUVB, in tension with certain observations, cannot solve the issue entirely.

\subsection{Comparing to Derived Gas Properties}
Just as \textsc{cloudy} can be used to predict column densities from simulated physical gas properties, measured column densities can be used to place constraints on the physical properties of the gas that is producing the absorption.  Here, we compare the observationally derived gas density and temperature of \citet{werk14} to those of the simulation.

\begin{figure}
\centering
\includegraphics[width=0.3\textwidth]{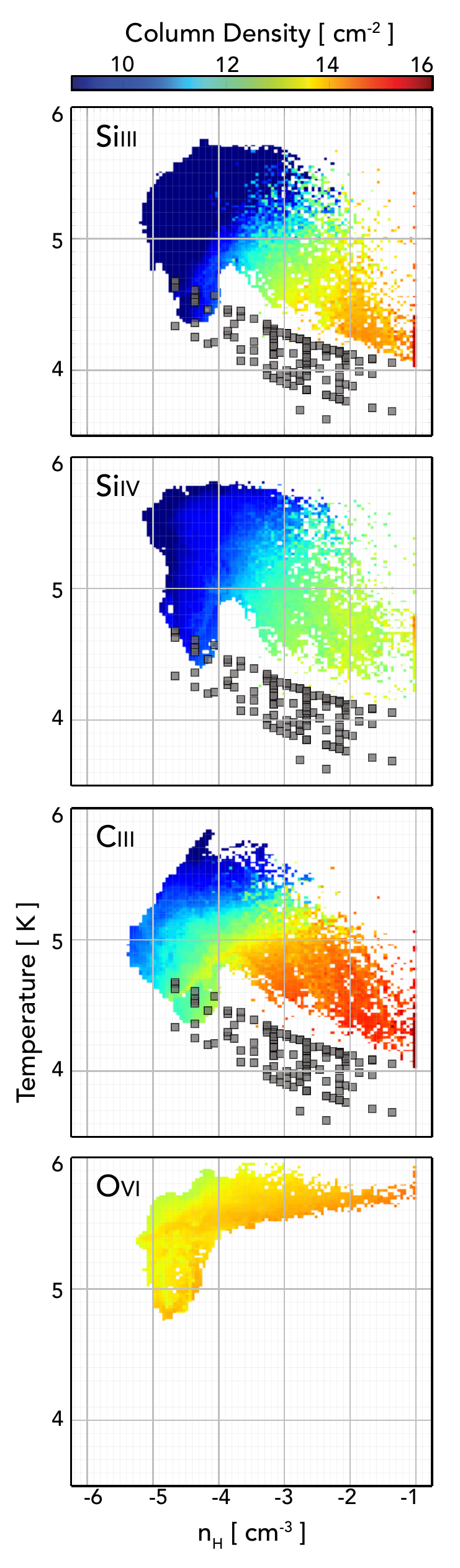}
\caption{Hydrogen number density ($n_{\mathrm{H}}$) and temperature, weighted by the given ion number density along the line of sight within the g1q1 model. Colors correspond to the average column density of lines of sight contributing to each bin.  Plotted square points are the values implied by the modeling of \citet{werk14}. The simulation and observations span the same range of densities while the simulation temperatures are universally higher.  This also shows the O{\scriptsize VI} clearly in a different phase medium while data points are not included as the ion is explicitly not fit by \citet{werk14}.  \label{hden_temp.fig}}
\end{figure}

We choose to compare the modeled temperature and density from the data directly to the simulation as opposed to computing mock spectra from the simulation and comparing to the data directly. Both require {\textsc cloudy} modeling and assumptions about the EUVB, abundance patterns and ionization equilibrium and in this way, we do not need to re-analyze the spectra of the COS Halos team.  Furthermore, the simulation densities and temperatures are weighted by the column density of the ion of interest, reflecting the preferential detection of higher column density features used in the modeling. 

For each of the column density radial profiles shown in Figure \ref{coldens_scatter.fig}, the hydrogen number density, $n_{\mathrm{H}}$, and temperature used in computing the column densities were projected along the same axis, weighted by the ion number density of interest within the g1q1 model. Figure \ref{hden_temp.fig} shows the resulting $n_{\mathrm{H}}$ and temperature for the four ions being discussed (Si{\scriptsize III}, Si{\scriptsize IV}, C{\scriptsize III}, O{\scriptsize VI}).  The 2D histogram is colored to show the average column density of the lines of sight contributing to each bin.  

The plotted points are derived from the modeling of the column density observations by \citet{werk14}.  The reported values of the ionization parameter, log(U), are directly converted to $n_{\mathrm{H}}$, assuming the ionizing flux of HM05.   For the temperature, \textsc{cloudy} models were generated using the adopted value of N$_{\mathrm{H}}$ and each permutation of the maximum and minimum values of log(U) and metallicity. For sight lines where there was an upper limit for the metallicity, a lower boundary of $[$Fe/H$] = -6$ was assumed.  In this way, each absorber has four data points associated with it, the combinations of the maximum and minimum values for log(U) and metallicity, representing the range of acceptable values from the data.  \textsc{cloudy} then reports the best equilibrium temperature of such a gas cloud, plotted here. This was done for the fiducial model, g1q1, which best matches the EUVB background used to derive the parameters in \citet{werk14}.

This plot more than any other shows that O{\scriptsize VI} is in a different phase from the other low-ionization species. It is found exclusively in the hottest gas and the fact that the column densities are uniform across the entire density range reflects the conditions necessary for its longer-lived existence instead of cooling immediately.  Data points are not shown for the modeled COS-Halos data as O{\scriptsize VI}  is explicitly left out of the \textsc{cloudy} modeling of \citet{werk14}, who are focused on cooler gas ($T < 10^5$ K).  

More surprising is the discrepancy between the properties of the low ions in the simulation and the observations.  For C{\scriptsize III}, Si{\scriptsize III}, and Si{\scriptsize IV}, the majority of the simulated points have log($n_{\mathrm{H}}) < -4$ but span a large range in temperature.  Conversely, the observational points have a clear relation where the temperature decreases with increasing density. This trend is perhaps reproduced in C{\scriptsize III} in the simulation but at higher temperatures. 

These plots show that gas with a measurable column density is found at low temperatures and high densities. Yet almost all of the gas in the simulation is at a higher temperature than those implied by \citet{werk14}. None of the gas within a radius of 100 kpc reaches this low of a temperature. Part of this may be due to observational selection; the majority of the simulated gas is at low density, below the detection threshold (see the second column of Figure \ref{hden_temp_evol.fig}). Alternatively, there may be physical differences between the simulated gas and the modeled observations that can explain the discrepancy in the column densities.  However, we have shown that this may also be alleviated by altering the assumed EUVB.

In short, although the simulated galaxy has column densities that can be brought into rough agreement with the data, the physical conditions of the gas producing such values is inconsistent with those derived from the data.  However, the differences between these two approaches should be noted.  \textsc{cloudy}, by design, constructs a cloud with uniform density and temperature in local ionization equilibrium. With only the EUVB fixed, it tends towards lower temperatures and produces a relationship between density and temperature seen as a track in the data plotted in Figure \ref{hden_temp.fig}.  The simulation, on the other hand, is run with the intent of retaining the large-scale and complex structure of the CGM gas, with ionization fractions computed for individual cells using a fixed temperature and density.  This work has shown that it is possible for this cosmological CGM to produce column densities in the range of those predicted by the idealized clouds intrinsic to \textsc{cloudy}, even with largely varying gas properties, if only the assumed EUVB is altered.

\section{Emission}
While the previous section demonstrates the power of absorption line studies, it also highlights some of their limitations. Absorption line measurements are extremely sensitive probes of low column density gas, but it is challenging to understand which physical structures in the halo we are probing with these data.  Also, the limited number of sightlines per galaxy hinders any attempt at understanding the spatial extent or scale of the structures detected. If instead it was possible to image the entire galaxy in emission, such maps may begin to reveal coherent structures or asymmetries in a gaseous halo that could help place stronger limits on gas accretion and outflows from the galaxy.  

Although the simulation fails to accurately predict the column density distribution of CGM gas, we feel that they have sufficient fidelity to obtain new estimates of CGM emission and to determine whether the emission signal is within reach of new observational capabilities. In particular, our analysis of CGM emission from a high resolution simulation of a galaxy at low redshift is distinct from most other recent studies that probe the overall, diffuse emission-line cooling from CGM halos at lower resolution.  This work provides a benchmark for simulations of this type for future comparisons with different theoretical prescriptions and observations. In this section, we present a complementary set of emission maps and show how they vary with photoionizing background and redshift.  We highlight the ways in which these maps provide a distinct and unique view of the morphology and evolution of the CGM. We also explore how emission varies with $n_{\mathrm{H}}$ and temperature in the context of the absorption discussed in the previous section.  We also determine the detectability of such gas, thereby informing target selection and instrument design for upcoming missions.

\subsection{Emission Maps}

\begin{figure}
\centering
\includegraphics[width=0.3\textwidth]{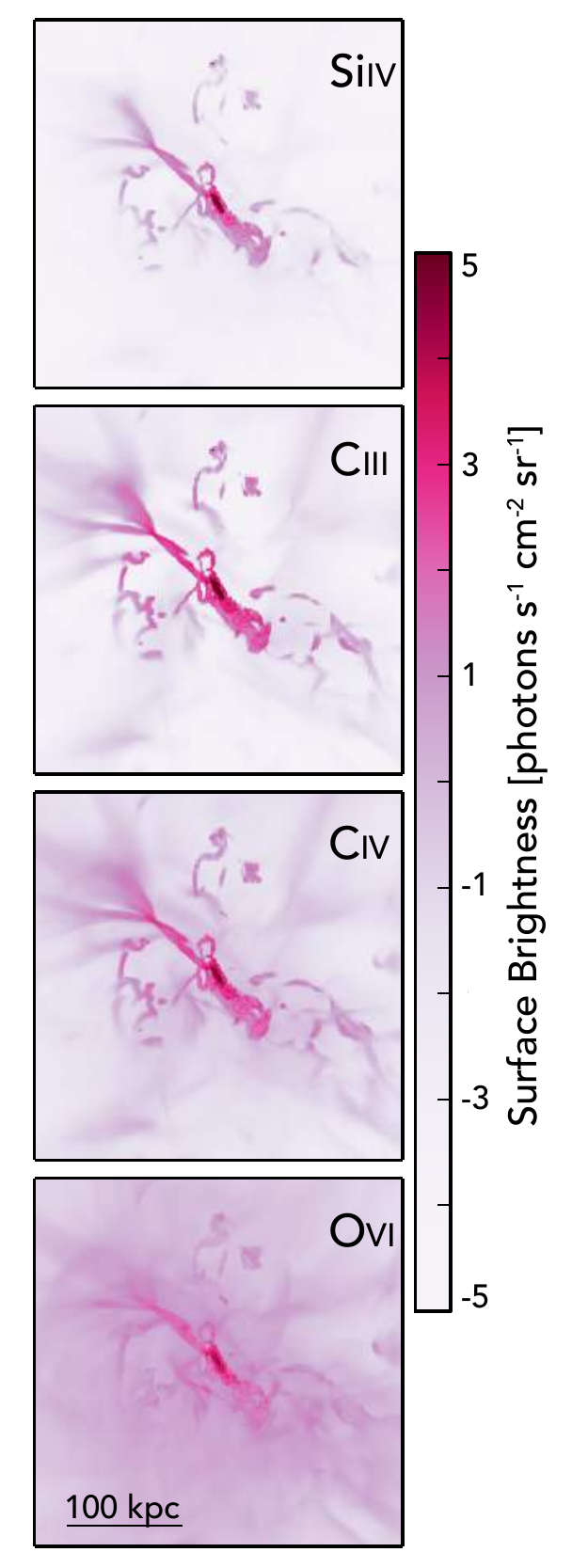}
\caption{Emission maps of Si{\scriptsize IV}, C{\scriptsize III}, C{\scriptsize IV} and O{\scriptsize VI} respectively at $z=0.2$ with a resolution of 1 kpc.  As in Figure \ref{abs_proj.fig}, O{\scriptsize VI} has the largest covering factor while the other ions more closely follow the underlying, cold gas structures. However, the surface brightness spans a wider range of values, demonstrating how sensitive the emission is to the density and temperature of the gas.  \label{emis_proj.fig}}
\end{figure}

Figure \ref{emis_proj.fig} shows the emission maps of four of the brightest lines at $z=0.2$, the same redshift as Figure \ref{abs_proj.fig} and with the same resolution.  Throughout this section, we consider C{\scriptsize IV} in place of Si{\scriptsize III} as it emits more brightly and is an intermediate ion between C{\scriptsize III} and O{\scriptsize VI}, except within roughly the innermost 25 kpc (a slightly smaller radius than that seen by \citet{vandeVoort_2013}). In general, Si{\scriptsize III} follows the same trends as Si{\scriptsize IV}, which is shown.  Furthermore, we do not present Lyman $\alpha$ emission (although it likely produces the strongest signal) because resonant scattering is known to change the extent and shape of the emission \citep[e.g.][]{lake2015,Dijkstra_2012,Zheng_2011} and the required radiative transfer calculation for Lyman $\alpha$ and other lines is beyond the scope of this paper and deferred for future work.  We assume that the impact of scattering for other emission lines is negligible. 

Comparing Figures \ref{abs_proj.fig} and \ref{emis_proj.fig}, it is visually apparent that emission and absorption trace the same high density structures. However, the emission surface brightness spans many more orders of magnitude making the relevant range much bigger than that of the column density measurements. A similarly large range was reported in earlier work by \citet{Bertone_2012}. Because much of the emission is expected to come from collisional excitation of the gas, the $n^2$ dependence of this process causes the large spread in values seen here and makes the higher density structures the brightest features and easiest to detect.  Most of the gas far from the disk is at low emission levels that are undetectable, as discussed below.  This suggests that even further from the galaxy, detections of the even lower density intergalactic medium will remain out of reach. 

Furthermore, as with the column densities, the low ions (Si{\scriptsize IV}, C{\scriptsize III}) trace the filaments where the density is higher and, just as importantly, the temperature is lower.  (The same is true for the not-shown Si{\scriptsize III} line, which is very similar to Si{\scriptsize IV}).  C{\scriptsize III} is the strongest emitting line, consistent with previous work that has considered the line \citep{vandeVoort_2013,Bertone_2013,Bertone_2012}.   As the peak of the emissivity curve for a given ion moves to higher temperatures, the emission becomes more prevalent for the majority of the diffuse halo which roughly has a temperature of ~$10^{5.5}$ K. For example, this is the peak temperature of O{\scriptsize VI} emissivity and as such, this line supplies strong emission throughout the entirety of the halo. Beyond ~$10^6$ K, however, the O{\scriptsize VI} emission will again become less volume filling as most of the CGM is not hotter than this. Such biasing of Si{\scriptsize IV} to higher density regions and O{\scriptsize VI} to less dense regions was also reported by \citet{vandeVoort_2013} and \citet{Bertone_2012} and appears to be a fundamental prediction of any simulation containing a warm/hot CGM halo. 

\begin{figure*}
\centering
\includegraphics[width=0.7\textwidth]{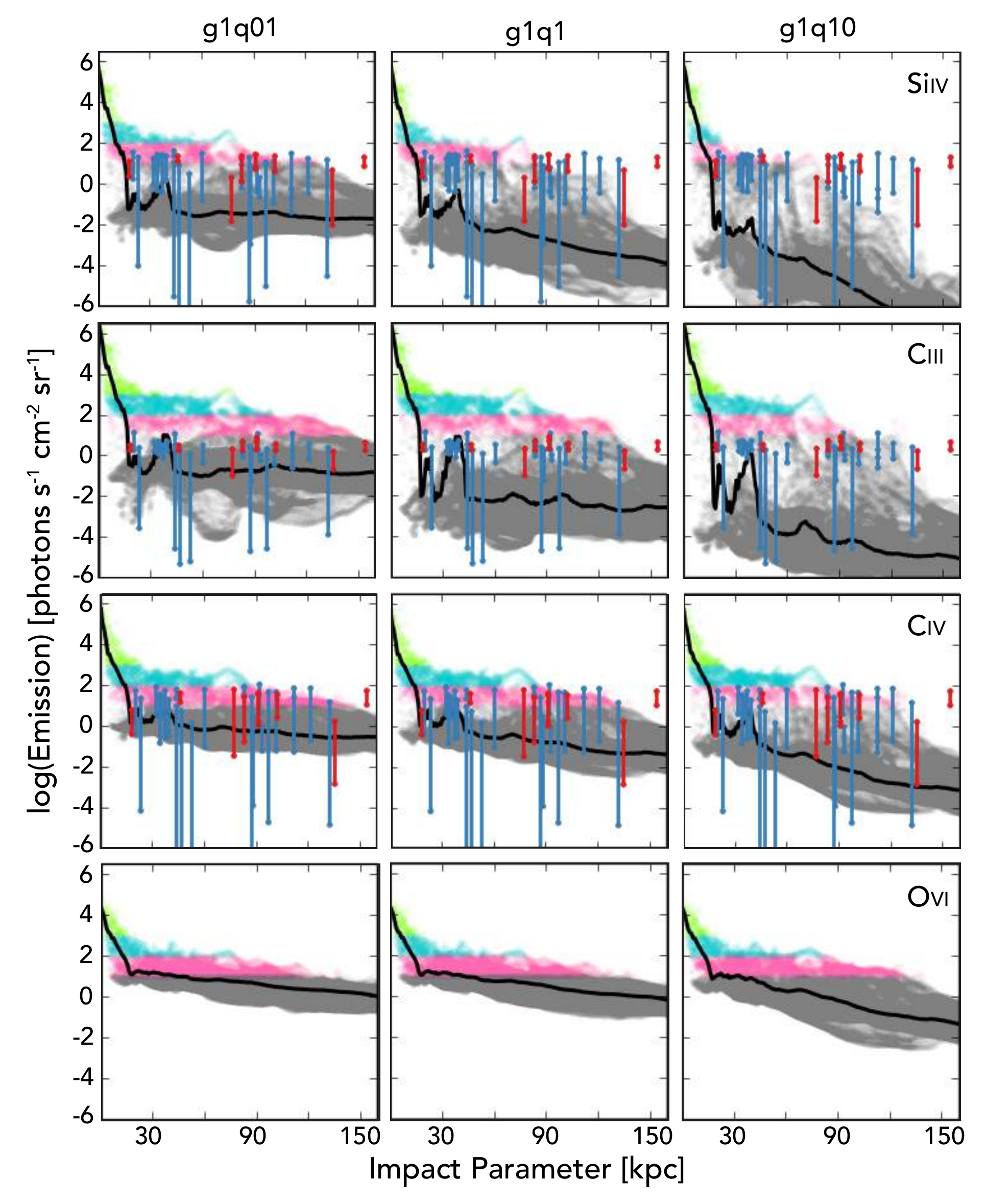}
\caption{Radial profiles of Si{\scriptsize IV}, C{\scriptsize III}, C{\scriptsize IV} and O{\scriptsize VI} emission at $z=0.2$ for the three EUVB backgrounds considered. Here the simulated points are colored by their possibility of detection with green being definite, blue being probable, and pink being possible. (See Section 4.4.1) The black line represents the median value. The median emission values are affected by the change in EUVB in the same way as the column densities (see Figure \ref{coldens_scatter.fig}). However, the extent and number of the brightest pixels with the highest detection probabilities are mostly unchanged by varying the EUVB.  Finally, overplotted are emission values associated with the density and temperature reported in \citet{werk14} for their absorbers. The large range for many of the points is a result of upper limits placed on the metallicity in the data.  Although the column densities were not in agreement, the emission values are in fact generally lower than what is predicted by the simulation.  \label{emis_scatter.fig}}
\end{figure*}

\subsection{Photoionizing Background}

The results above assume the single, fiducial HM05 ionizing background.  It was shown in the previous section that varying the intensity of this background can bring the simulation more in-line with the absorption observations but what effect does this have on the predicted emission? Figure \ref{emis_scatter.fig} shows the same radial profiles as Figure \ref{coldens_scatter.fig} but for the surface brightness values of each projected pixel.  A comparison of these two plots shows that the median values of the emission is affected by the change in EUVB in mostly the same way as the column densities. To further quantify this notion, Table \ref{abs_emis_100kpc.tab} shows the median value of the radial profiles of both the column density and surface brightness predictions at 100 kpc and $z=0.2$ for the three EUVBs considered in both sections.  For Si{\scriptsize IV} and C{\scriptsize III}, the median values of the column density increase by almost two orders of magnitude in the g1q01 model relative to the fiducial model and the same is true for the emission.  The O{\scriptsize VI} is more unaffected both in absorption and emission. When considering g1q10 relative to g1q1, both the column density and surface brightness medians decrease by the same orders of magnitude for Si{\scriptsize IV} and C{\scriptsize III}, and the O{\scriptsize VI} by 0.6 orders of magnitude. The ionizing background now over-ionizes the low-density gas as before, leading to larger changes in both absorption and emission of all ions, including O{\scriptsize VI}.

However, these median levels of emission are below the detection limits of upcoming surveys for all the models and thus the EUVB does not have a strong effect on what will realistically be possible to detect.  In Figure \ref{emis_scatter.fig}, the points are colored by rough detection probability cuts: green points are certain to be detected, blue points are likely to be detected and pink points are possible to detect. This coloring shows that the pixels with the highest surface brightnesses extend about equally far for all of the EUVBs.  Instead, their distribution and number depends mostly on the density and temperature of the gas. Because a 2D image would capture all of these pixels, the number and brightness of these emission peaks can perhaps provide a more unambiguous look at these underlying gas properties.  Such surface brightness effects are discussed in more detail in Section 4.4.1.

More importantly however, the pixels with the highest surface brightness have the highest possibility of detection and extend about equally far for all of the EUVBs. Their distribution and number depends mostly on the density and temperature of the gas. Because a 2D image would capture all of these pixels, the number and brightness of these emission peaks can perhaps provide a more unambiguous look at these underlying gas properties.

\begin{table}[h]
\caption{Median Values of z=0.2 Radial Profiles at 100 kpc \label{abs_emis_100kpc.tab}}
\begin{center}
\begin{tabular}{ c c c c } 
\multicolumn{4}{c}{\textsc{absorption}}  \\ \hline
 & g1q01 & g1q1 & g1q10  \\ \hline 
 Si{\scriptsize IV} & 11.46 & 9.45 & 7.00     \\ 
 C{\scriptsize III} & 13.18 & 11.22 & 8.76  \\ 
 O{\scriptsize VI} & 13.75 & 13.55 & 12.66 \\ 
 \hline
 \multicolumn{4}{c}{log(Column Density)}  \\
 \multicolumn{4}{c}{  [cm$^{-2}$] }  \\ 
 \\ 

\multicolumn{4}{c}{\textsc{emission}}  \\ \hline
 & g1q01 & g1q1 & g1q10 \\ \hline 
 Si{\scriptsize IV} & -1.41  & -2.84 & -5.24   \\ 
 C{\scriptsize III} & -0.47 & -1.96 & -4.23 \\ 
 O{\scriptsize VI} &  0.50 & 0.33 & -0.46 \\ 
\hline
 \multicolumn{4}{c}{log(Surface Brightness)} \\
\multicolumn{4}{c}{ [photons s$^{-1}$ cm$^{-2}$ sr$^{-1}$ ] }  \\ 
\end{tabular}
\end{center}
\end{table}

This variation of the emission radial profile with EUVB is opposed to previous studies which found no variation in their simulated emission profiles when the assumed background was increased by a factor of ten although neither explore lowering the EUVB \citep{Sravan_2015,vandeVoort_2013}.  Also, both average over a large number of halos over range of masses, which may smooth some of the changes seen here, particularly as we are presenting the median.  However, creating the same radial profiles as Figure \ref{emis_scatter.fig} but for $z=1$, we also found little to no variation in the profiles. Yet the fiducial background is higher at $z=1$ than at $z=0.2$ and encompasses the regions covered by the g1q1 and g1q10 models at $z=0.2$.  This suggests that the gas state is more dominated by collisional ionization at $z=1$, which is supported by the higher density of the gas at early times (see Section 4.3 and Figure \ref{hden_temp_evol.fig} for more details).

We point out that the extremely low SB values predicted here are purely theoretical predictions from the gas itself. They do not include the EUVB photons as well as possible photon pumping and scattering from the host galaxy continuum.  The UV continuum of a star forming galaxy typical of the COS Halos sample will be on the order of 500 photon s$^{-1}$ cm$^{-2}$ sr$^{-1}$, although this is dependent on the uncertain escape fraction of the ionizing photons \citep[see Figure 13 of ][for a plotted, typical galaxy SED]{werk14}. With only a fraction of this flux contributing to the pumping or scattering of a specific line of interest, the floor set by these processes will be above the theoretical limit shown here but below any upcoming detectable limit. Furthermore, this will mostly affect the volume closer to the star-forming disk and not alter our predictions for the more distant CGM. 

In addition to the projected simulation pixels, ``observational'' data has been generated from the physical parameters inferred from the \textsc{cloudy} modeling of \citet{werk14}. For each line of sight, the adopted N$_{\mathrm{H}}$ was paired with each combination of the maximum and minimum values of the metallicity and ionization parameter to compute the emission from such a cloud using \textsc{cloudy}. Then the maximum and minimum computed values of each sight line are plotted as connected points. The large acceptable range for a high faction of the sightlines is due to the degeneracy of the ionization parameter (which for this model is a proxy for $n_{\mathrm{H}}$) and the metallicity.  As we previously noted,  the measured column densities are not in agreement with the simulated values.  However the emission predicted from the column density data using the method described above, bracket nearly all of the simulated emission values of Si{\scriptsize III}, Si{\scriptsize IV}, and C{\scriptsize III}.  Data points are not shown for O{\scriptsize VI} since \citet{werk14} explicitly model gas cooler than the gas seen in the simulation ($T < 10^5$ K).  

The complex ionization structure in the CGM halo means that regions of strong absorption do not always produce significant associated emission, particularly for O{\scriptsize VI} which is known to have a low column. In part, this is because absorption-line measurements are typically probing gas in the ground state, allowing ionic absorption of incoming quasar photons. Emission, on the other hand, is generated as higher ions cool through the metal line of interest or through collisional excitation and cooling of lower ion gas; these processes tend to be more transitory in nature.  We suggest here that the C{\scriptsize III} emission predictions may be the most reliable as the column density distribution is best reproduced by the simulation both for the fiducial EUVB and the modified, weaker EUVB. The column density of O{\scriptsize VI} is known to be underproduced and the temperature of the hot halo generating the O{\scriptsize VI} is contested amongst the simulations (discussed further in Section 5). However, we expect the overall trends seen here with O{\scriptsize VI} being found in the hotter, volume-filling gas to remain valid.  C{\scriptsize IV}  is likely intermediary but the lack of data in the COS Halos sample limits our conclusions.   

Overall, the density structures revealed in the column density maps are also present in the emission but the emission values span a much larger dynamic range. This dynamic range reflects the emission's biasing towards higher density and thus higher signal regions. However, this biasing and it's unique dependence on density, temperature and metallicity can provide complementary constraints on these properties of the gas when combined with column density measurements.  Another advantage of emission observations is that varying the ionizing background does not have a strong effect on the brightest emitting pixels as it does on the median column density that could be detected.  Because the emission trends and detection possibilities do not vary with the EUVB, in the following sections, we present results using the fiducial background. 

\subsection{Redshift Evolution of the Emission}
Our simulations can also be used to predict the physical distribution of CGM emission over a a large range of redshift and imaging the CGM can be used to better understand what is driving the emission at each redshift. In particular, the nature of the CGM may change dramatically from $z=1$ to $z=0$ as the SFR declines on average and as galaxies in this mass range potentially transition from ``cold mode'' to ``hot mode'' accretion. 

\begin{figure*}
\centering
\includegraphics[width=1.0\textwidth]{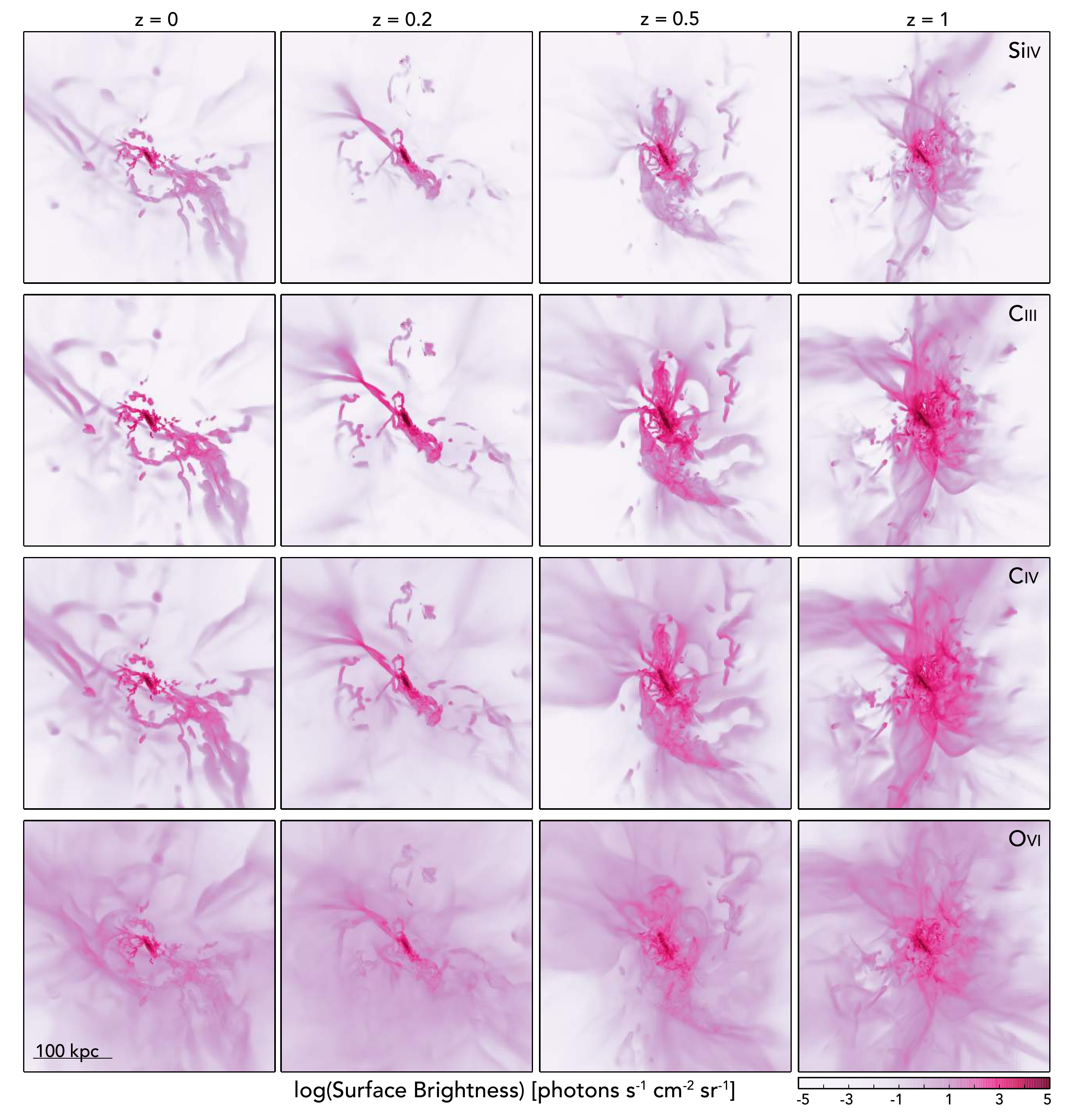}
\caption{Emission maps for Si{\scriptsize IV}, C{\scriptsize III}, C{\scriptsize IV}, and O{\scriptsize VI} at $z=0, 0.2, 0.5, 1.0$ with a resolution of 1 physical kpc and with a fiducial HM05 background showing that the intrinsic brightness increases with redshift (that is the brightness without accounting for the $(1+z)^4$ dimming - i.e. as if $z=0$ for the distance). The emission becomes more filamentary with lower redshift as the increasing gas temperature and decreasing average density shift the brightest emission to these remaining high density regions. \label{emis_theory_zevol.fig}}
\end{figure*}

\begin{figure}
\centering
\includegraphics[width=0.475\textwidth]{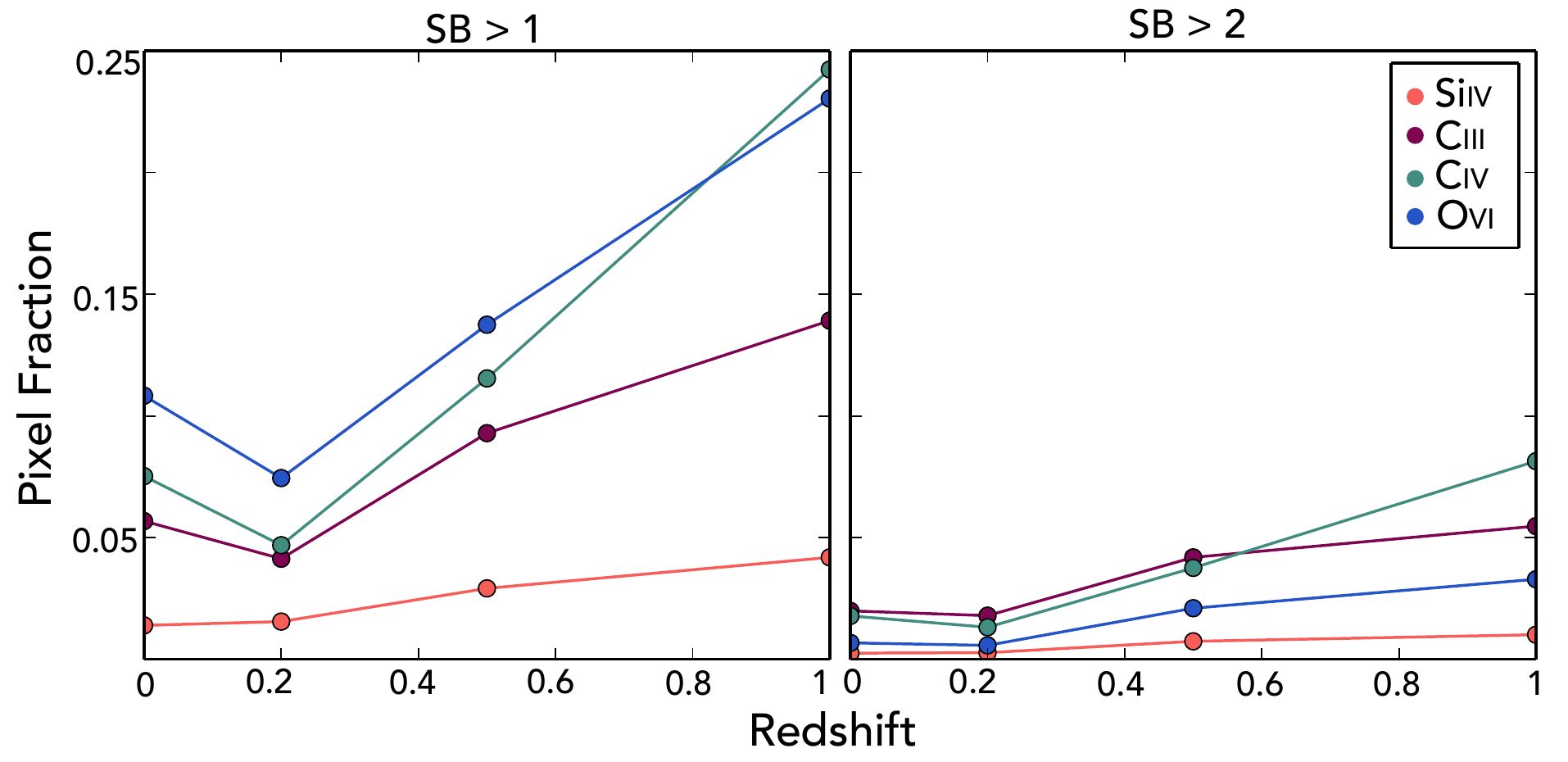}
\caption{Covering fraction for the 4 lines of interest at two different surface brightness cutoffs: 10 and $10^2$ photons s$^{-1}$ cm$^{-2}$ sr$^{-1}$ respectively. This excludes the galactic disk.  In general, the emission increases with redshift for all lines. At the lower level, O{\scriptsize VI} has the highest covering fraction at all redshifts but the highest one. At the higher surface brightness level, C{\scriptsize III} is the dominant ion except at $z=1$ where C{\scriptsize IV} increases rapidly. This shows the overall increasing emission with increasing redshift, the high covering fraction of low-SB O{\scriptsize VI}, and that the strongest emission is coming from ions with mid-ionization energies.  (Here again the $(1+z)^4$ dimming is not accounted for - i.e. as if $z=0$ but this has no effect on the trends.)  \label{cover_frac.fig}}
\end{figure}

Figure \ref{emis_theory_zevol.fig} shows the evolution of four emission lines for our simulated galaxy using the fiducial background. The proper physical size of the box is constant in each panel as well as the resolution of 1 physical kpc.  Furthermore, in order to study the intrinsic evolution of the emission separate from the cosmic expansion, these plots show the true surface brightness of the object without accounting for the cosmological $(1+z)^4$ dimming - i.e. we set $z=0$. In this way, the brightness of the emission is directly related to the underlying density, temperature and metallicity and their evolution alone. We discuss the importance and effects of this dimming in the following section. 

It is easy to see that as the redshift increases, the emission becomes brighter and extends further and more spherically from the disk. This is most striking for the low ions which emit at an appreciable level almost out to the virial radius at $z=1$ while they are limited to high-density features at $z=0$.  However, the extended emission sphere surrounding the disk at $z=1$ in all four ions considered disappears by $z=0$, leaving only the filamentary structure behind. This increase in the bias of the emission towards high density regions at later times as compared to emission at $z=3$ was noted by \citet{vandeVoort_2013}, and we point out that this is true even when comparing $z=1$ to the present day.  

These changes in the emission values are more easily seen when quantified as a covering fraction.  Figure \ref{cover_frac.fig} shows the evolution of the fraction of pixels with an intrinsic emission above two different surface brightness limits [10 and 100 photons s$^{-1}$ cm$^{-2}$ sr$^{-1}$] for the four ions of interest within a square image that is 320 proper kpc per side with a resolution of 1 proper kpc. Disk pixels have been removed to emphasize the CGM.   As redshift increases, the increasing brightness seen in Figure \ref{emis_theory_zevol.fig} leads to higher covering fractions for the emission.   For both limits, the fraction doubles between $z=0$ and $z=1$ for all the ions. O{\scriptsize VI} has the largest covering fraction at the lowest level considered  except at $z=1$.  C{\scriptsize III} and C{\scriptsize IV} have similar covering fractions for all the limits and are the dominant lines at the higher SB cutoff, seen in the right pane of Figure \ref{cover_frac.fig}. Most interestingly, by $z=1$, C{\scriptsize IV} has the largest covering fraction, overtaking C{\scriptsize III} and O{\scriptsize VI}.  Even at the lowest surface brightness limit that we consider, the covering fraction is never higher than 0.25 for any ion as early as $z=1$, highlighting the difficulty of emission detections.

\begin{figure*}
\centering
\includegraphics[width=0.8\textwidth]{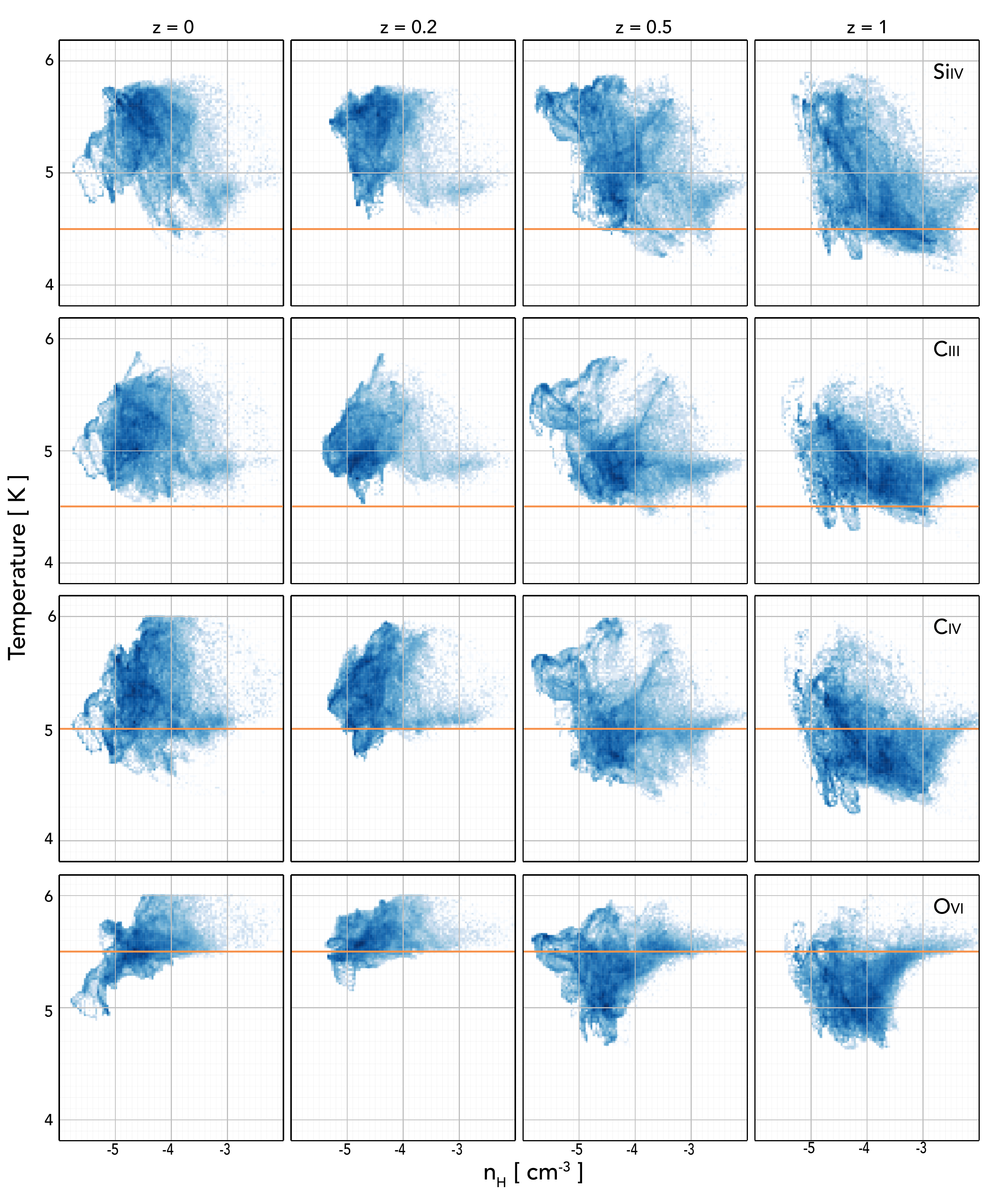}
\caption{Evolution with redshift of the hydrogen number density ($n_{\mathrm{H}}$) versus temperature, weighted by the emissivity of a given line. The orange lines show the temperature of the peak of the emissivity curve for that ion. For all ions, the trend is for the gas to move to lower densities and higher temperatures as the galaxy evolves from $z=1$ to $z=0$. This change in the gas represents the decreasing role of the filaments feeding the galaxy as well as the cumulative effect of SN-driven outflows.  \label{hden_temp_evol.fig}}
\end{figure*}

To better understand these changes in the surface brightness maps, Figure \ref{hden_temp_evol.fig} shows the density temperature diagram of the galaxy weighted by the emissivity of multiple ions for four different redshifts ($z=0,0.2,0.5,1.0$).  For each of the ions, there is a clear trend towards lower densities and higher temperatures on average with decreasing redshift.  The decrease in density causes a lower overall surface brightness across the majority of the halo.  Simultaneously, the higher temperatures move the bulk of the gas away from the peak of the emissivity curve of the lower ions.  In this way, the emission from most of the volume is reduced, leaving only the higher density and lower temperature filaments with an appreciable signal. Once again, O{\scriptsize VI} defies these trends as the higher temperatures at late times are more in line with its emissivity peak, accounting for its continued higher surface brightness throughout the area shown in Figure \ref{emis_theory_zevol.fig}.

\begin{figure}
\centering
\includegraphics[width=0.5\textwidth]{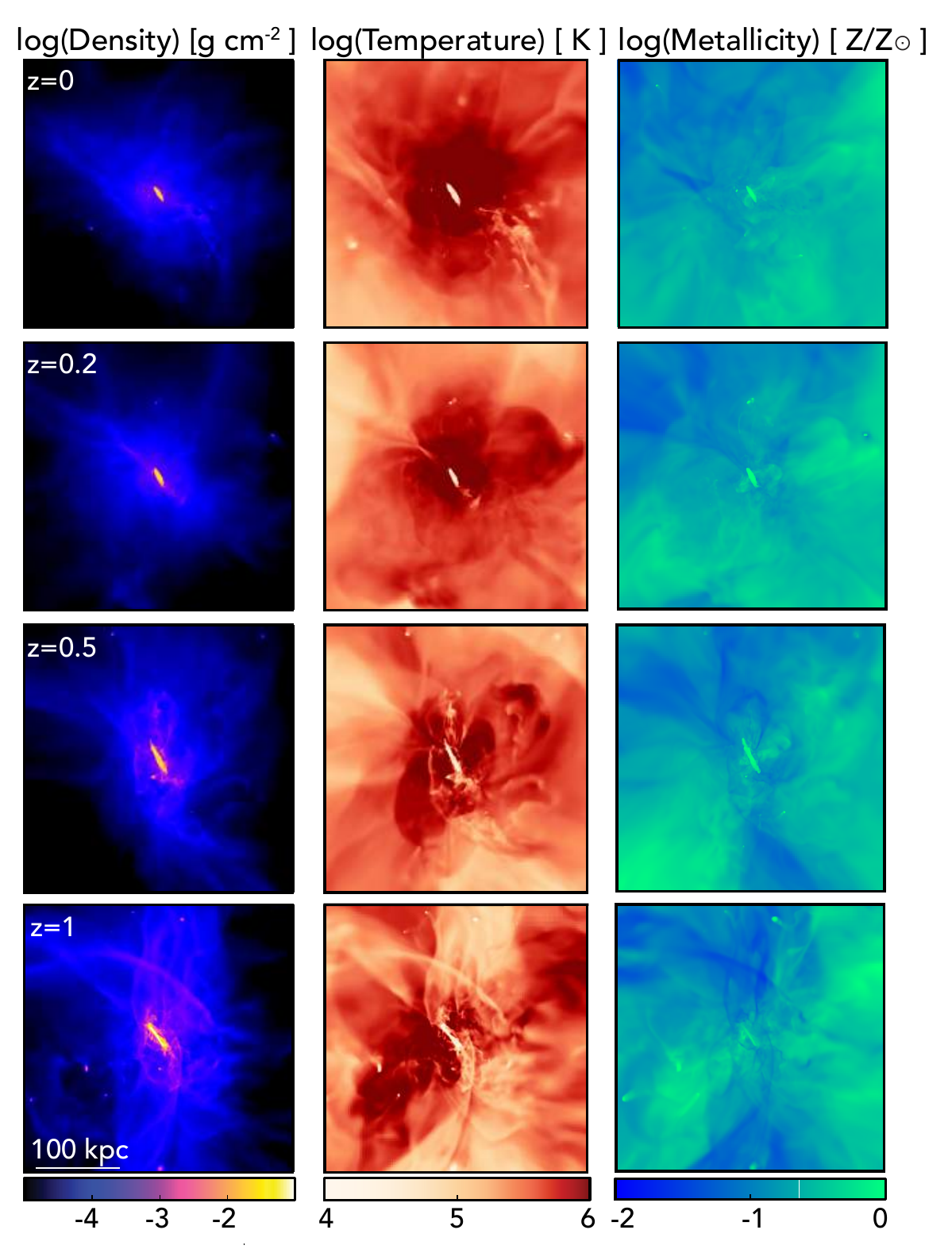}
\caption{Projections of the density (left) and the density-weighted temperature (middle) and metallicity (right). The combined evolution of these quantities is what drives the changes predicted for the emission. Filaments are easily seen feeding the galaxy at $z=1$ in the density and as low metallicity regions and have weakened by $z=0$. The temperatures become higher and more uniform by $z=0$.     \label{dens_temp_projs.fig}}
\end{figure}

The question then becomes: what is causing these systematic changes in the density and temperature of the CGM? While this is difficult to answer definitively, there are two dominant effects within the simulation: accretion and supernova feedback.  The first effect arises from filaments feeding the galaxy, as seen in the density projections in the left column of Figure \ref{dens_temp_projs.fig}, showing the density evolution of the galaxy and its CGM.  At $z=1$, there are three well-defined filaments penetrating the galactic halo down to the disk. As the redshift decreases, the galaxy mass increases, and cosmic expansion lowers the overall average density of the IGM, these features become broader and do not penetrate into the halo as deeply although they do supply additional gas along with stripped satellite material \citep{ryan, ximena}.  Instead, the gas density profile becomes more spherical and more extended as the galaxy evolves. 

What's surprising is that this change in CGM morphology is not reflected in the structure of the brighter emission. The fractured filaments exist as surface brightness peaks in the $z=1$ projection but the bright emission halo is more spherical whereas the brighter emission at $z=0$ is almost entirely contained in the remains of the filaments with no discernible symmetry.

\begin{figure*}
\centering
\includegraphics[width=0.75\textwidth]{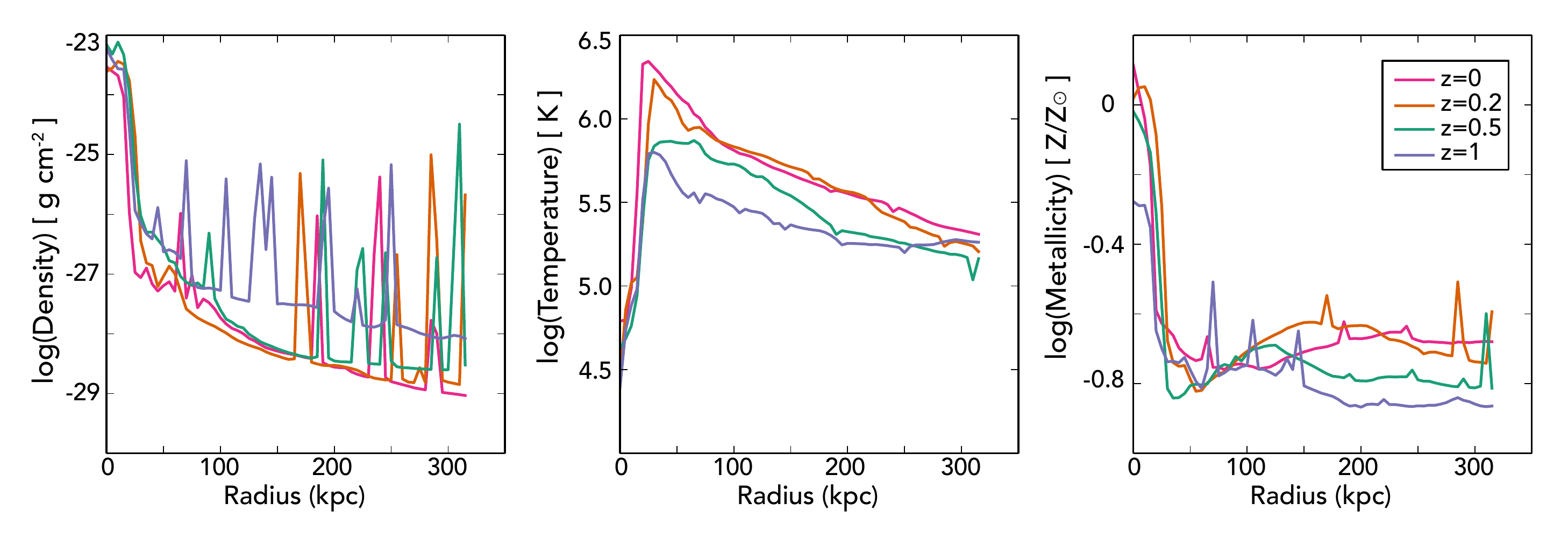}
\caption{Radial profiles of density, temperature, and metallicity at a series of redshifts. The spherical averages of these quantities are plotted. They quantitatively demonstrate the trends driving the evolution of the emission - the average density lowers with time while the temperature increases. Spikes in the profiles correspond to substructure in the halo of the main galaxy.    \label{radial_profiles.fig}}
\end{figure*}

The explanation resides in the corresponding temperature projections, shown in the middle column of Figure \ref{dens_temp_projs.fig}.  At $z=1$, the halo has a complex temperature structure, with colder, denser gas contained in cold, in-falling gas and satellites.  By $z=0$, this is replaced with a spherical hot halo, mimicking the density profile.  Even the filaments are bringing in predominantly warm/hot gas \citep{ryan}. The weakening of the filaments, the slowing of their supply of cold gas, and their replacement with a more uniformly hot halo of lower density gas results in the loss of the large emission halos of $z=1$.  By $z=0$, only the remaining cold, dense features are capable of producing a significant signal. Figure \ref{radial_profiles.fig} shows the radial profiles of the relevant quantities at various redshifts, which quantitatively demonstrates these trends.

In addition to this change in accretion mode, supernova-driven winds are also effective at creating low-density, high-temperature pressurized bubbles in their wake as they expand through the halo. At high redshift, it's been suggested that galactic outflows are the dominant process powering the time-varying simulated emission \citep{Sravan_2015}.  Looking again at the temperature projections of Figure \ref{dens_temp_projs.fig}, higher temperature regions extend perpendicular to the disk. Furthermore, multiple spherical plumes can be seen expanding away from the disk.  These features are less visible in the density but still seen. This suggests that outside the filament regions, SN-driven outflows play a large role in shaping the density and temperature distribution of the halo.  In particular, because the SN energy is injected as thermal feedback in this simulation, the temperature of the gas is efficiently raised to $10^{5.5}$K on average, much higher than the peak emissivity of low ion lines, such as C{\scriptsize III} and C{\scriptsize IV}.

In conclusion, at later times, the emission in the metal lines studied here is more structured, tracing the remaining high density structures, in contrast to the majority of the gas which becomes more spherically distributed and more uniformly hot. This emission may be an effective way to probe continued galactic accretion at low redshift. Additionally, the propensity for the gas to become hotter and more diffuse translates to a decrease in the magnitude and extent of the low-ion emission. Galactic winds coupled with a transition to hot mode accretion no longer resupplying cold gas likely explains the shift to low-density, high-temperature gas at late times.  

\subsection{Implications for Detection}
Finally, with an emission signal this faint, understanding how these theoretical predictions relate to what can actually be detected is important for both furthering interpretations of future measurements as well as enabling fair comparisons of theory and observations.  In this section, we examine how realistic surface brightness and angular resolution limitations can limit the conclusions that can be drawn about a galaxy's CGM.

\subsubsection{Surface Brightness Limits}
Recent work in CCD technology now allow us to reach extremely high quantum efficiencies in the UV \citep{Hamden_2012} leading to achieving unprecedented low surface brightness limits in the UV. In this section, we look to see how this translates into reaching levels where UV emission from the CGM of nearby galaxies can finally be detected. 

We consider three regimes of detection possibility for the emission. Pixels with a surface brightness (SB) greater than $10^3$ photons s$^{-1}$ cm$^{-2}$ sr$^{-1}$ are certain to be detected by the specifications of any upcoming instrument and are colored green in the following plots. Those with $10^2 <$ SB $< 10^3$ photons s$^{-1}$ cm$^{-2}$ sr$^{-1}$ have a high probability of being detected and are plotted in blue. Finally, pixels with $10 <$ SB $< 10^2$ photons s$^{-1}$ cm$^{-2}$ sr$^{-1}$ have a possibility of being detected and are shown in pink. Exact confidence levels will vary for a given instrument and observing strategy but these are appropriate rules of thumb. 

\begin{figure*}
\centering
\includegraphics[width=1.0\textwidth]{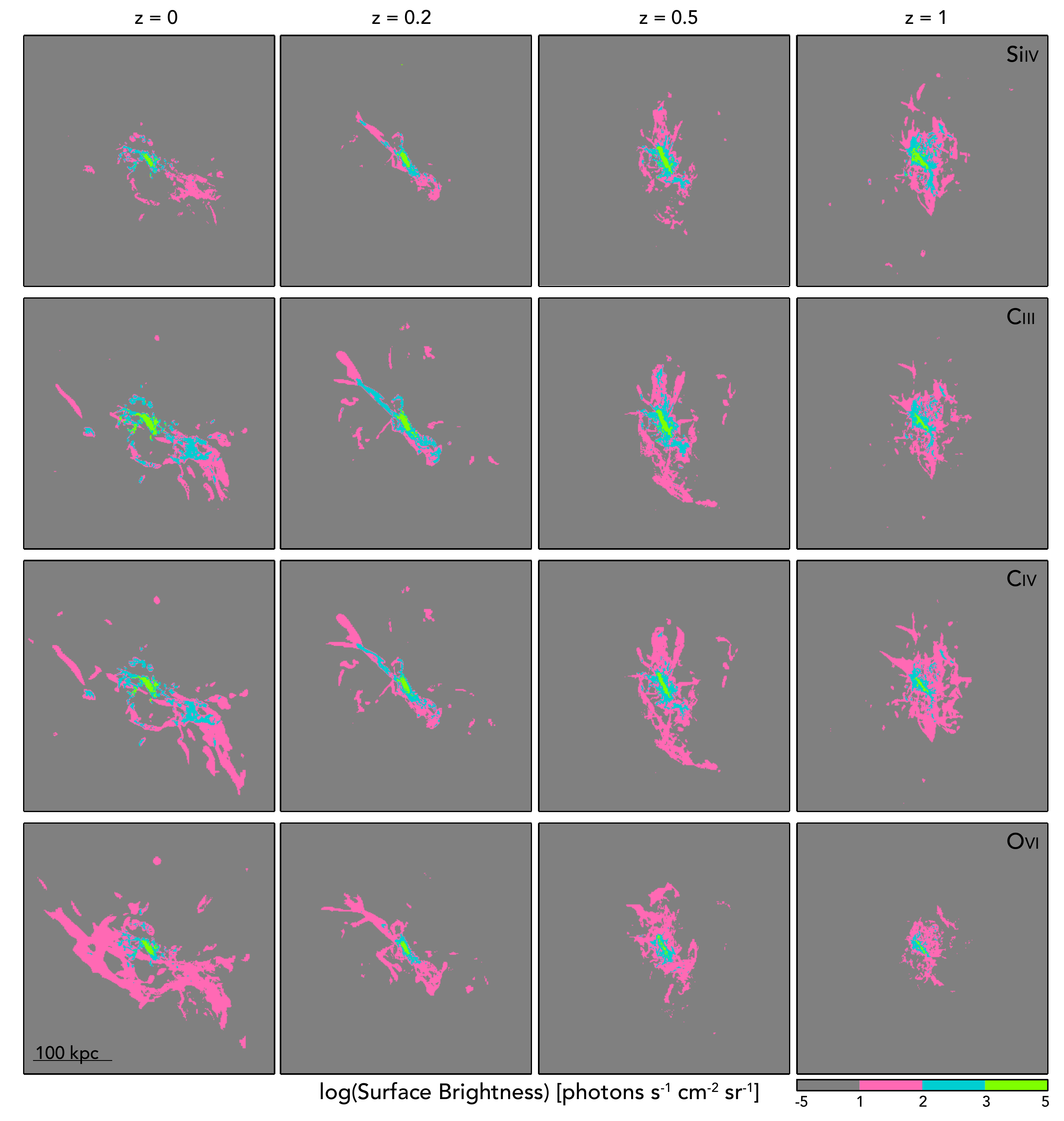}
\caption{Surface brightness maps for three different ions at $z=0,0.2, 0.5,1$.  Each map takes into account the surface brightness dimming due to the given redshift. Like in the theoretical predictions of Figure \ref{emis_theory_zevol.fig}, the emission becomes more filamentary with time. However, because of the intrinsic brightening of the gas, the extent of the emission varies little between $z=0.5$ and $z=1$. O{\scriptsize VI} is affected the most because it emits primarily at the lowest detectable range at $z=0$ which is then dimmed from detection.  \label{emis_varyZ.fig}}
\end{figure*}

Figure \ref{emis_varyZ.fig} again shows the surface brightness maps of Figure \ref{emis_theory_zevol.fig} but now colored to show these detection probabilities.  Unlike the theoretical projections, these maps take into consideration the $(1+z)^4$ dimming of the surface brightness due to the expansion of the universe.  The brightest, easiest to detect emission (green) always comes from the galactic disk. Some pixels reach this brightness level into the filaments, especially in C{\scriptsize III}, but only at the lower redshifts. 

Much more promising for CGM studies is the extent of the blue pixels, indicating regions that are likely to be detected. For the brightest ions (C{\scriptsize III}, C{\scriptsize IV},), these regions extend into a large portion of the filaments at $z=0$, out to as far as 100 kpc. The mid-level emission is less extended for the high ion, O{\scriptsize VI}, reaching a radius of only 50 kpc \citep[similar to that found in previous work by ][]{vandeVoort_2013,Furlanetto_2004}. Thus, although Figure \ref{emis_varyZ.fig} shows that the extent of the emission decreases with redshift, the detectable emission is still an appreciable distance from the main galactic disk.  At all redshifts it should be possible to detect emission from CGM gas beyond the galactic (star-forming) disk.

Finally, the lowest surface brightness limits naturally reveal the most extended structure and emphasize the importance of pushing the limits of future instrumentation.  However, in considering a range of redshift, the combination of the surface-brightness dimming and the fixed surface brightness limits shapes the observable covering fraction.  The radial profiles of Figure \ref{emis_scatter.fig} suggest how this is possible, where the colors of the simulated points correspond to the same limits as the projections of Figure \ref{emis_varyZ.fig}. At $z=0.2$, most of the points lie below any reasonable detection limit and the fraction at the lowest detection limit (pink) is greater than at the higher limits (blue, green) at all but the smallest radii. This trend is true at all redshifts. For O{\scriptsize VI} in particular, much of the emission is intrinsically emitted at the lowest limit considered here because it is mostly generated by the low-density, volume-filling gas. At $z=1$, most of this dim emission then falls below observational levels once the cosmological dimming is considered.  In this way, the extent of possible O{\scriptsize VI} detections drops from 130 kpc at $z=0$ to 60 kpc at $z=1$.  

For the low ions, the effect is less pronounced because they emit most brightly and significantly in the relatively over-dense filaments and this emission in fact increases with increasing redshift as seen in Figure \ref{emis_theory_zevol.fig} \citep[demonstrated also in][]{Bertone_2013}.  The surface brightness dimming is thus offset by the inherent increased emissivity, seen as an increased theoretical covering fraction in Figure \ref{cover_frac.fig}. Thus, the decrease in detection extent is less steep yet still pronounced for C{\scriptsize III} (150 kpc to 100 kpc) and C{\scriptsize IV} (150 kpc to 90 kpc).  Furthermore, most of this decrease is in place by $z=0.5$ and little change in the observable properties of the gas happens between $z=0.5$ and $z=1$. Work at higher redshifts indicates that these extents are decreasing slightly in physical scales but that they remain at the same fraction of the halo's virial radius \citep{Sravan_2015, vandeVoort_2013}.

Thus, in order to make a clear detection of the CGM, it is necessary to reach a detection limit of at least 100 photons s$^{-1}$ cm$^{-2}$ sr$^{-1}$ to begin to probe extended emission in dense filamentary regions. Pushing down to 10 photons s$^{-1}$ cm$^{-2}$ sr$^{-1}$ provides the possibility of detecting emission from the diffuse, hot volume-filling phase of the CGM.  Furthermore, observations close to $z=0$ increase the chances of detecting this phase as it is the first to drop out of range due to surface brightness dimming. However, between $z=0.5$ and $z=1$, the extent of the emission is relatively unchanged at these detection limits, enabling measurements across redshift that could trace the evolution of the CGM.   

\subsubsection{Resolution Limitations}
In addition to the surface brightness limits, the resolution of the image affects the types of conclusions that can be drawn from the observations. For all of the previous plots, the resolution of the projected grid has been set to 1 kpc physical such that the angular resolution varies with redshift. At this resolution, it is possible to see the filaments and streams feeding the galaxy. How does this change if the physical resolution is varied?

\begin{figure}
\centering
\includegraphics[width=0.475\textwidth]{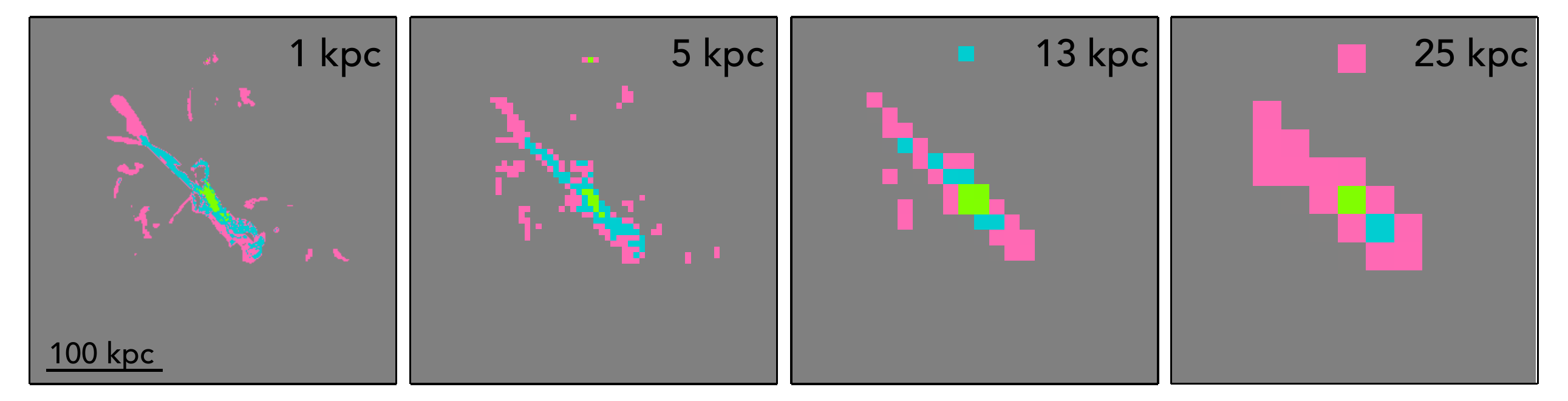}
\caption{C{\scriptsize III} emission at $z=0.2$ for four different resolutions - the fiducial 1 kpc, 5 kpc, 13kpc, and 25 kpc proper corresponding to angular resolutions of 0.3'', 1.5'', 4" and 7.6'' respectively. The medium resolutions reproduce many of the features of the highest resolution and would allow for a more confident detection of filamentary CGM emission features. At the lowest resolution, it is possible to detect an elongation of the emission and the distances would make it identifiable as CGM material but filaments are less conclusive. \label{res_vary.fig}}
\end{figure}

Figure \ref{res_vary.fig} shows the C{\scriptsize III} emission for the galaxy at $z=0.2$, with four different resolutions: the fiducial 1 kpc, 5 kpc, and 25 kpc.  This corresponds to angular resolutions of 0.3'', 1.5'', 4" and 7.6'' respectively.  At the moderate resolution of 13 kpc, it is still possible to discern the filamentary features extending from the disk. At the lowest resolution, however, the CGM emission resembles an extended halo around the galaxy. Because of its physical extent, one can still associate this emission with the CGM but valuable information about the spatial distribution of the gas has been lost.  The low resolution also makes it difficult to track the evolution of the CGM. At most, an elongation aligned with the disk could be seen developing, corresponding to the filaments feeding the disk but internal clumpy structures and features perpendicular to the disk are masked.

One optimistic consequence of these predictions derives from the fact that at higher redshift, the physical angular diameter of an object is almost unchanging due to cosmological expansion.  Given the relatively constant physical extent of the emission, the resolution of any observation will not change much for galaxies between $z=0.5$ and $z=1$. In this range, the 25 kpc physical resolution corresponds to roughly 3''-4'' and the 5 kpc resolution to 0.6''-0.8''.  Thus, by ensuring the resolution is at least 4'', it should be possible to confirm emission from the CGM for $0.5<z<1$ and begin to resolve its structure at lower redshifts.  At $z=0.2$, basic filamentary structure and stripped material should be distinguishable with this angular resolution. (See the 13 kpc panel of Figure \ref{res_vary.fig}.  Within the next year, the balloon-borne FIREBall-2 will launch with this resolution and is expected to make a positive detection.  A small, UV emission line explorer is currently being designed to have a comparable resolution but also to cover a much larger range in wavelength, providing complementary coverage to FIREBall-2.  Finally, further in the future, a 12-meter class, UV/optical space telescope named the High-Definition Space Telescope (HDST) has been designed with the specification of 0.01'' angular resolution between 100-500 nm. \footnote{The report describing this proposed telescope can be found at \url{http://www.hdstvision.org/}} As shown in Figure \ref{res_vary.fig}, this unprecedented resolution would allow for an evaluation of the predominance of filamentary accretion for the first time as well as structures created by galactic outflows.

\section{Discussion}
It is only recently that comparing the CGM of simulations to data has become possible and that emission-line predictions from the simulations are relevant for upcoming observations. We are entering a new realm of detailed CGM studies for which it is necessary to understand the limitations of current simulations and to evaluate which conclusions we expect to remain robust. 

In this paper, we have examined a single simulation of a Milky Way-like galaxy using one form of purely thermal supernova feedback in an {\tt enzo} AMR simulation. Like \citet{hummels} who uses a similar prescription, we find that the simulation has difficulty in producing the necessary column densities for all the ions but especially O{\scriptsize VI}. \citet{hummels} found that implementing stronger feedback brought better agreement while we've found that lowering the assumed EUVB can be equally effective.  On the contrary, the most common way to implement supernova feedback in SPH simulations is to give either a constant or physically-scaled velocity kick to a series of wind particles which carry away the SN energy in kinetic form. One such simulation by \citet{ford} found better agreement with the low-ions although they still fail to reproduce the O{\scriptsize VI} observations.  This implies that the direct, thermal methods prevalent in AMR causes the gas to reach higher temperatures as opposed to the wind velocity approach of SPH simulations, which shock heat differently.  Furthermore, the simple thermal feedback assumed in this simulation also leads to the well known over-cooling problem. The H{\scriptsize I} distribution of the galaxy is known to be too centrally concentrated at z=0 \citep{ximena} and too many stars are formed \citep{ryan}.

However, new methods are emerging that incorporate additional components of the SN feedback. For example, \citet{Liang_2015} included prescriptions for supernova pressure and momentum in addition to a thermal heating model in a series of RAMSES AMR simulations. Their fiducial model was also not a good fit to low-$z$ data and again, increasing the feedback and lowering the star formation efficiency led to greater agreement.  Building on the wind velocity method, \textsc{arepo} and Illustris in particular include an implementation of AGN feedback with quasar and radio modes. \citet{Suresh_2015b} found that including the radio mode in particular is responsible for enriching the CGM in the simulations and reproduces the bimodality of star-forming and passive galaxies seen in O{\scriptsize VI} data. However, they still see a stronger mass dependence on the O{\scriptsize VI} distribution than what is observed and the radio mode feedback in Illustris is known to be too extreme, removing too much gas from the center of massive galaxies \citep{Suresh_2015b,genel_2014}.  \citet{oppenheimer_2016} found that AGN feedback in their EAGLE SPH simulations does not have a large effect on the OVI column densities and that the lower values for passive galaxies is instead driven by their higher virial temperature.  Additionally, the inclusion of non-equilibrium chemistry in their simulations also does not resolve the almost universal issue of  producing too little O{\scriptsize VI} for star-forming galaxies and passive galaxies alike.  Finally, \citet{Salem_2015} found that implementing a two-fluid cosmic ray method resulted in a cosmic ray-driven wind that gradually accelerated the gas, allowing for a larger range of gas temperatures as well as a higher metallicity beyond 100 kpc. Both contribute to higher column densities for all of the ions considered in this paper. In particular, the simulation reproduced the O{\scriptsize VI} measurements of the COS Halos survey for star forming galaxies.

In short, reproducing both the stellar properties and the CGM properties of a given galaxy at low redshift is a major theoretical challenge and an important test of modern simulation methods. The majority have difficulties in capturing the multiphase medium required to produce such high levels of low and high ions in the data. Likely, a combination of these advanced feedback prescriptions will be necessary to remedy this. Thus, low-$z$ CGM absorption measurements are a powerful new way to constrain such prescriptions and further motivates the emission observations we are predicting in this work. Emission predictions can provide complementary constraints on these feedback processes. Understanding the role and effect of purely thermal supernova feedback in work such as this will allow us to estimate its importance in future work with more complex schemes. 

In addition to the uncertainty in the feedback scheme, the resolution of the simulation potentially limits the conclusions that can be drawn from comparisons with the column density data.   \citet{werk14}  found low number densities for the cool clouds causing the absorption in the COS Halos data, corresponding to cloud sizes of 0.1-2000 pc, the larger of which can be resolved by current zoom-in simulations. However, \citet{crighton} detected the presence of 8 smaller ($<$100-500 pc), higher density clouds in an QSO spectrum with a z=2.5 foreground galaxy.  If these scales are the norm, simulations may not resolve these clouds which exist beyond the high density disk and which potentially contain a large fraction of the cool CGM gas. These resolution concerns and the possible existence of dense clouds relate back to the feedback mechanisms required to accurately reproduce the multiphase medium. Dense clumps will have cool, self-shielding cores that could explain the observations of low ions while the shells around them and the diffuse volume-filling gas could be responsible for the mid to high ions. Instead, lower resolution simulations may produce a mid-range, average temperature that does not precisely reflect the state of the gas.  Further studies of the cooling of the gas in idealized simulations where the resolution can be much higher could tell us more about how cool gas forms and persists within the volume-filling hot phase of the galactic halo. In addition, if dense small clouds do exist, they should appear as bright points in emission studies as opposed to lower density clouds since the emission scales as the square of the density, offering a chance to probe the number and density of the emitting clouds if the angular resolution is high enough.  

Finally, this is ultimately a simulation of a single Milky Way-like galaxy. The conclusions drawn here about the filamentary structure of the gas especially at low redshift might be specific to this particular galaxy.  More than just changing the viewing angle is necessary for understanding the dependence of our predictions on the physical properties of the gas. Our exact expectations could change if the galaxy forms fewer stars; if the filaments are broader than expected here; if the environment of the galaxy increases the metallicity of the gas - to name a few examples.  The cosmological study of emission sources by \citet{Frank_2012} suggest that an appreciable number of sources will be detectable at the redshifts considered in this paper such that we can begin to measure how the emission varies with these physical parameters.  Thus, it is only by conducting both future observational surveys and a larger range of cosmological simulations that we can begin to address the variance in emission signatures of the CGM. 

Even with these limitations of the simulation, we expect the trends seen in our conclusions to be robust.  Varying the EUVB can produce variations in the simulated column densities at low redshift and in future work this can be further explored in addition to alternative forms of feedback.  Furthermore, there is seemingly a tension between the density and temperature of gas within simulations of this type and those modeled from the COS Halos measurements.  The coexistence of large amounts of Si{\scriptsize IV}, C{\scriptsize III} and O{\scriptsize VI} and our failure to reproduce all three simultaneously indicates that there is a \emph{more} complex temperature structure than what's seen here. Because of this, details of the extent and shape of the emission may vary in future work but the dominance of the C{\scriptsize III}, C{\scriptsize IV}, and O{\scriptsize VI} lines has remained thus far and should persist.  Similarly, low ions tracing higher density structures while high ions are more volume filling is a clear prediction of any warm/hot gaseous halo. In addition, altering feedback methods to capture a larger range of temperatures could lead to further structure in the emission at low redshift. Less clear is how numerous and how thin these features might become. In this way, the resolution of future observations could be the limiting factor in imaging the CGM structures, possibly even more so than surface brightness limits. However, the general conclusions of this paper regarding the required surface brightness limits and resolution limits should continue to reflect simulations of this type.

\section{Summary and Conclusions}
Observing the predicted gas halos of nearby galaxies has long been a goal of observations but the need to study this gas in the space ultraviolet coupled with the diffuse nature of the gas in question has made this challenging.  Now, studies of the CGM at low redshift are entering an unprecedented age of sensitivity.  Measurements can begin to constrain theoretical prescriptions in simulations as well as discriminate between them.  In this work, we vary the EUVB in a high-resolution cosmological simulation of a Milky Way-like galaxy and examine its role in determining how the simulated column densities compare to recent data. We then predict the emission signal expected from such gas for upcoming instrumentation as well as how it varies with redshift and the physical properties of the gas itself. 

Our main conclusions can be listed as follows:
\begin{enumerate}
\item Looking at column density maps at $z=0.2$, the largest values for the column densities of all the ions studied here are found in high density filamentary structures. The low-ions (H{\scriptsize I}, C{\scriptsize III}, Si{\scriptsize IV}) are found almost exclusively in these structures while O{\scriptsize VI} is found throughout the halo as its higher ionization energy allows it to exist in the volume-filling hot gas.
\item Varying the quasar component of the standard EUVB can significantly change the predicted column densities of the simulation. In particular, lowering this component by a factor of 100 brings the simulation values into much better agreement with the low-ion data of the COS Halos sample.  The simulated O{\scriptsize VI} column densities remain too low at all impact parameters compared to the observed values for star-forming galaxies, even with the strongest EUVB. 
\item Comparing the gas temperature and density in the simulation to that found through \textsc{cloudy} modeling of the COS Halos data shows that the simulation predicts higher temperatures than the data modeling. This demonstrates that it is possible to produce similar column densities from different gas distributions.
\item Examining the redshift evolution of the emission reveals that the emission becomes more structured at later times, tracing the remaining high density, low temperature features. This is in contrast to the majority of the gas which shifts to lower densities and higher temperatures from $z=1$ to $z=0$ due to the weakening of cold gas filaments and the progression of supernova-driven winds. 
\item A surface brightness limit of 100 photons s$^{-1}$ cm$^{-2}$ sr$^{-1}$ should enable a clear detection of emission from the CGM with, C{\scriptsize III} emission extending as far as 100 kpc and O{\scriptsize VI} as far as 50 kpc at $z=0.2$. The predicted extent stays roughly constant for $0.5 < z< 1.0$ as the cosmological surface brightness dimming is balanced by an increasing intrinsic emissivity.
\item An angular resolution of 4'' is necessary to begin to resolve the spatial distribution of the CGM out to $z=1$ and sub-arcsecond resolution is needed to resolve beyond a general elongation from the disk.  At $z=0.2$, this same observations require an angular resolution of 7.6" (for elongation) and 1.5" (for features) respectively.
\end{enumerate}

These conclusions focus on results from the combination of predicted UV absorption and emission-line data from a simulated Milky Way-like galaxy, offering a physical explanation for the trends seen in observations for the existence and extent of multiple ions. Other studies have focused on varying feedback prescriptions to bring simulations into better agreement with recent data. However, this can also be reversed as simulation predictions can be extended to create true mock observations that can enable better interpretations of future data. To make more accurate predictions for observations, future work will have to include a number of details excluded here.

First, the low surface brightness of the emission in question means that the UV background can overpower the CGM signal. Including a model of the background signal and incorporating its subtraction will provide a better understanding of which CGM structures can be detected with confidence.  Second, the continuum emission from the galaxy can also dominate the CGM emission-line signal close to the disk, especially at moderate to low resolution. The disk-halo interface is where the SN winds are being launched; understanding this transition is particularly important.  Finally, the velocity structure of the gas has not been considered here, which can change the line profiles of the emission.  \citet{ryan} examined the flow of gas into and out of the galaxy, finding that the majority of the accretion at low redshift was in the form of warm/hot gas.  Associating emission with these flows is left for future work but will become crucial as integral field units that provide both spatial and spectral information are becoming commonplace. This kinematic information will provide the best observational evidence for both inflows and outflows of gas from galaxies. 

We are grateful to M. Ryan Joung for generously sharing his simulation output and guidance. L.C. would also like to thank Yuan Li, Cameron Hummels, Munier Salem and Bruno Milliard for helpful discussions. L.C. and D.S. acknowledge support from NASA grant NNX12AF29G. L.C. would also like to acknowledge support from the Chateaubriand Fellowship.  

\bibliographystyle{apj}    
\bibliography{library}

\begin{thebibliography}{}
\expandafter\ifx\csname natexlab\endcsname\relax\def\natexlab#1{#1}\fi

\bibitem[{{Adams} {et~al.}(2011){Adams}, {Uson}, {Hill}, \&
  {MacQueen}}]{adams_2011}
{Adams}, J.~J., {Uson}, J.~M., {Hill}, G.~J., \& {MacQueen}, P.~J. 2011, \apj,
  728, 107

\bibitem[{{Agertz} {et~al.}(2013){Agertz}, {Kravtsov}, {Leitner}, \&
  {Gnedin}}]{Agertz_2013}
{Agertz}, O., {Kravtsov}, A.~V., {Leitner}, S.~N., \& {Gnedin}, N.~Y. 2013,
  \apj, 770, 25

\bibitem[{{Agertz} {et~al.}(2011){Agertz}, {Teyssier}, \&
  {Moore}}]{Agertz_2011}
{Agertz}, O., {Teyssier}, R., \& {Moore}, B. 2011, \mnras, 410, 1391

\bibitem[{{Arrigoni Battaia} {et~al.}(2015){Arrigoni Battaia}, {Yang},
  {Hennawi}, {Prochaska}, {Matsuda}, {Yamada}, \& {Hayashino}}]{Battaia_2015}
{Arrigoni Battaia}, F., {Yang}, Y., {Hennawi}, J.~F., {et~al.} 2015, \apj, 804,
  26

\bibitem[{{Barai} {et~al.}(2013){Barai}, {Viel}, {Borgani}, {Tescari},
  {Tornatore}, {Dolag}, {Killedar}, {Monaco}, {D'Odorico}, \&
  {Cristiani}}]{Barai_2013}
{Barai}, P., {Viel}, M., {Borgani}, S., {et~al.} 2013, \mnras, 430, 3213

\bibitem[{{Benson} {et~al.}(2013){Benson}, {Venkatesan}, \&
  {Shull}}]{benson_escape}
{Benson}, A., {Venkatesan}, A., \& {Shull}, J.~M. 2013, \apj, 770, 76

\bibitem[{{Bertone} {et~al.}(2013){Bertone}, {Aguirre}, \&
  {Schaye}}]{Bertone_2013}
{Bertone}, S., {Aguirre}, A., \& {Schaye}, J. 2013, \mnras, 430, 3292

\bibitem[{{Bertone} \& {Schaye}(2012)}]{Bertone_2012}
{Bertone}, S., \& {Schaye}, J. 2012, \mnras, 419, 780

\bibitem[{{Booth} {et~al.}(2013){Booth}, {Agertz}, {Kravtsov}, \&
  {Gnedin}}]{Booth_2013}
{Booth}, C.~M., {Agertz}, O., {Kravtsov}, A.~V., \& {Gnedin}, N.~Y. 2013,
  \apjl, 777, L16

\bibitem[{{Booth} \& {Schaye}(2009)}]{Booth_2009}
{Booth}, C.~M., \& {Schaye}, J. 2009, \mnras, 398, 53

\bibitem[{{Bridge} {et~al.}(2013){Bridge}, {Blain}, {Borys}, {Petty},
  {Benford}, {Eisenhardt}, {Farrah}, {Griffith}, {Jarrett}, {Lonsdale},
  {Stanford}, {Stern}, {Tsai}, {Wright}, \& {Wu}}]{Bridge_2013}
{Bridge}, C.~R., {Blain}, A., {Borys}, C.~J.~K., {et~al.} 2013, \apj, 769, 91

\bibitem[{{Brook} {et~al.}(2011){Brook}, {Governato}, {Ro{\v s}kar}, {Stinson},
  {Brooks}, {Wadsley}, {Quinn}, {Gibson}, {Snaith}, {Pilkington}, {House}, \&
  {Pontzen}}]{Brook_2011}
{Brook}, C.~B., {Governato}, F., {Ro{\v s}kar}, R., {et~al.} 2011, \mnras, 415,
  1051

\bibitem[{{Brooks} {et~al.}(2009){Brooks}, {Governato}, {Quinn}, {Brook}, \&
  {Wadsley}}]{Brooks_2009}
{Brooks}, A.~M., {Governato}, F., {Quinn}, T., {Brook}, C.~B., \& {Wadsley}, J.
  2009, \apj, 694, 396

\bibitem[{{Bryan} {et~al.}(2014){Bryan}, {Norman}, {O'Shea}, {Abel}, {Wise},
  {Turk}, {Reynolds}, {Collins}, {Wang}, {Skillman}, {Smith}, {Harkness},
  {Bordner}, {Kim}, {Kuhlen}, {Xu}, {Goldbaum}, {Hummels}, {Kritsuk}, {Tasker},
  {Skory}, {Simpson}, {Hahn}, {Oishi}, {So}, {Zhao}, {Cen}, {Li}, \& {Enzo
  Collaboration}}]{enzo}
{Bryan}, G.~L., {Norman}, M.~L., {O'Shea}, B.~W., {et~al.} 2014, \apjs, 211, 19

\bibitem[{{Cen} {et~al.}(2005){Cen}, {Nagamine}, \& {Ostriker}}]{Cen_2005}
{Cen}, R., {Nagamine}, K., \& {Ostriker}, J.~P. 2005, \apj, 635, 86

\bibitem[{{Ceverino} {et~al.}(2014){Ceverino}, {Klypin}, {Klimek},
  {Trujillo-Gomez}, {Churchill}, {Primack}, \& {Dekel}}]{Ceverino_2014}
{Ceverino}, D., {Klypin}, A., {Klimek}, E.~S., {et~al.} 2014, \mnras, 442, 1545

\bibitem[{{Chabrier}(2003)}]{Chabrier_2003}
{Chabrier}, G. 2003, \pasp, 115, 763

\bibitem[{{Chen} {et~al.}(2010){Chen}, {Helsby}, {Gauthier}, {Shectman},
  {Thompson}, \& {Tinker}}]{Chen_2010}
{Chen}, H.-W., {Helsby}, J.~E., {Gauthier}, J.-R., {et~al.} 2010, \apj, 714,
  1521

\bibitem[{{Cooray}(2016)}]{cooray_2016}
{Cooray}, A. 2016, ArXiv e-prints, arXiv:1602.03512

\bibitem[{{Crighton} {et~al.}(2015){Crighton}, {Hennawi}, {Simcoe}, {Cooksey},
  {Murphy}, {Fumagalli}, {Prochaska}, \& {Shanks}}]{crighton}
{Crighton}, N.~H.~M., {Hennawi}, J.~F., {Simcoe}, R.~A., {et~al.} 2015, \mnras,
  446, 18

\bibitem[{{Dav{\'e}} {et~al.}(2013){Dav{\'e}}, {Katz}, {Oppenheimer},
  {Kollmeier}, \& {Weinberg}}]{Dave_2013}
{Dav{\'e}}, R., {Katz}, N., {Oppenheimer}, B.~D., {Kollmeier}, J.~A., \&
  {Weinberg}, D.~H. 2013, \mnras, 434, 2645

\bibitem[{{Dav{\'e}} {et~al.}(2011){Dav{\'e}}, {Oppenheimer}, \&
  {Finlator}}]{Dave_2011}
{Dav{\'e}}, R., {Oppenheimer}, B.~D., \& {Finlator}, K. 2011, \mnras, 415, 11

\bibitem[{{Dijkstra} \& {Kramer}(2012)}]{Dijkstra_2012}
{Dijkstra}, M., \& {Kramer}, R. 2012, \mnras, 424, 1672

\bibitem[{{Dove} {et~al.}(2000){Dove}, {Shull}, \& {Ferrara}}]{dove_escape}
{Dove}, J.~B., {Shull}, J.~M., \& {Ferrara}, A. 2000, \apj, 531, 846

\bibitem[{{Ferland} {et~al.}(1998){Ferland}, {Korista}, {Verner}, {Ferguson},
  {Kingdon}, \& {Verner}}]{cloudy}
{Ferland}, G.~J., {Korista}, K.~T., {Verner}, D.~A., {et~al.} 1998, \pasp, 110,
  761

\bibitem[{{Fern{\'a}ndez} {et~al.}(2012){Fern{\'a}ndez}, {Joung}, \&
  {Putman}}]{ximena}
{Fern{\'a}ndez}, X., {Joung}, M.~R., \& {Putman}, M.~E. 2012, \apj, 749, 181

\bibitem[{{Ford} {et~al.}(2013){Ford}, {Oppenheimer}, {Dav{\'e}}, {Katz},
  {Kollmeier}, \& {Weinberg}}]{Ford_2013}
{Ford}, A.~B., {Oppenheimer}, B.~D., {Dav{\'e}}, R., {et~al.} 2013, \mnras,
  432, 89

\bibitem[{{Ford} {et~al.}(2015){Ford}, {Werk}, {Dave}, {Tumlinson}, {Bordoloi},
  {Katz}, {Kollmeier}, {Oppenheimer}, {Peeples}, {Prochaska}, \&
  {Weinberg}}]{ford}
{Ford}, A.~B., {Werk}, J.~K., {Dave}, R., {et~al.} 2015, ArXiv e-prints,
  arXiv:1503.02084

\bibitem[{{Frank} {et~al.}(2012){Frank}, {Rasera}, {Vibert}, {Milliard},
  {Popping}, {Blaizot}, {Courty}, {Deharveng}, {P{\'e}roux}, {Teyssier}, \&
  {Martin}}]{Frank_2012}
{Frank}, S., {Rasera}, Y., {Vibert}, D., {et~al.} 2012, \mnras, 420, 1731

\bibitem[{{Fumagalli} {et~al.}(2011){Fumagalli}, {Prochaska}, {Kasen}, {Dekel},
  {Ceverino}, \& {Primack}}]{Fumagalli_2011}
{Fumagalli}, M., {Prochaska}, J.~X., {Kasen}, D., {et~al.} 2011, \mnras, 418,
  1796

\bibitem[{{Furlanetto} {et~al.}(2004){Furlanetto}, {Schaye}, {Springel}, \&
  {Hernquist}}]{Furlanetto_2004}
{Furlanetto}, S.~R., {Schaye}, J., {Springel}, V., \& {Hernquist}, L. 2004,
  \apj, 606, 221

\bibitem[{{Gawiser} {et~al.}(2007){Gawiser}, {Francke}, {Lai}, {Schawinski},
  {Gronwall}, {Ciardullo}, {Quadri}, {Orsi}, {Barrientos}, {Blanc}, {Fazio},
  {Feldmeier}, {Huang}, {Infante}, {Lira}, {Padilla}, {Taylor}, {Treister},
  {Urry}, {van Dokkum}, \& {Virani}}]{Gawiser_2007}
{Gawiser}, E., {Francke}, H., {Lai}, K., {et~al.} 2007, \apj, 671, 278

\bibitem[{{Genel} {et~al.}(2014){Genel}, {Vogelsberger}, {Springel}, {Sijacki},
  {Nelson}, {Snyder}, {Rodriguez-Gomez}, {Torrey}, \& {Hernquist}}]{genel_2014}
{Genel}, S., {Vogelsberger}, M., {Springel}, V., {et~al.} 2014, \mnras, 445,
  175

\bibitem[{{Governato} {et~al.}(2007){Governato}, {Willman}, {Mayer}, {Brooks},
  {Stinson}, {Valenzuela}, {Wadsley}, \& {Quinn}}]{Governato_2007}
{Governato}, F., {Willman}, B., {Mayer}, L., {et~al.} 2007, \mnras, 374, 1479

\bibitem[{{Guo} {et~al.}(2010){Guo}, {White}, {Li}, \&
  {Boylan-Kolchin}}]{guo_2010}
{Guo}, Q., {White}, S., {Li}, C., \& {Boylan-Kolchin}, M. 2010, \mnras, 404,
  1111

\bibitem[{{Haardt} \& {Madau}(1996)}]{HM96}
{Haardt}, F., \& {Madau}, P. 1996, \apj, 461, 20

\bibitem[{{Haardt} \& {Madau}(2001)}]{HM01}
{Haardt}, F., \& {Madau}, P. 2001, in Clusters of Galaxies and the High
  Redshift Universe Observed in X-rays, ed. D.~M. {Neumann} \& J.~T.~V. {Tran},
  64

\bibitem[{{Haardt} \& {Madau}(2012)}]{HM12}
---. 2012, \apj, 746, 125

\bibitem[{{Hamden} {et~al.}(2012){Hamden}, {Greer}, {Schiminovich}, {Nikzad},
  \& {Martin}}]{Hamden_2012}
{Hamden}, E.~T., {Greer}, F., {Schiminovich}, D., {Nikzad}, S., \& {Martin},
  D.~C. 2012, in Society of Photo-Optical Instrumentation Engineers (SPIE)
  Conference Series, Vol. 8453, Society of Photo-Optical Instrumentation
  Engineers (SPIE) Conference Series, 9

\bibitem[{{Hopkins} {et~al.}(2014){Hopkins}, {Kere{\v s}}, {O{\~n}orbe},
  {Faucher-Gigu{\`e}re}, {Quataert}, {Murray}, \& {Bullock}}]{hopkins_fire}
{Hopkins}, P.~F., {Kere{\v s}}, D., {O{\~n}orbe}, J., {et~al.} 2014, \mnras,
  445, 581

\bibitem[{{Hopkins} {et~al.}(2012){Hopkins}, {Quataert}, \&
  {Murray}}]{Hopkins_2012}
{Hopkins}, P.~F., {Quataert}, E., \& {Murray}, N. 2012, \mnras, 421, 3522

\bibitem[{{Hummels} \& {Bryan}(2012)}]{Hummels_2012}
{Hummels}, C.~B., \& {Bryan}, G.~L. 2012, \apj, 749, 140

\bibitem[{{Hummels} {et~al.}(2013){Hummels}, {Bryan}, {Smith}, \&
  {Turk}}]{hummels}
{Hummels}, C.~B., {Bryan}, G.~L., {Smith}, B.~D., \& {Turk}, M.~J. 2013,
  \mnras, 430, 1548

\bibitem[{{Joung} {et~al.}(2012){Joung}, {Putman}, {Bryan}, {Fern{\'a}ndez}, \&
  {Peek}}]{ryan}
{Joung}, M.~R., {Putman}, M.~E., {Bryan}, G.~L., {Fern{\'a}ndez}, X., \&
  {Peek}, J.~E.~G. 2012, \apj, 759, 137

\bibitem[{{Kere{\v s}} {et~al.}(2005){Kere{\v s}}, {Katz}, {Weinberg}, \&
  {Dav{\'e}}}]{Keres_2005}
{Kere{\v s}}, D., {Katz}, N., {Weinberg}, D.~H., \& {Dav{\'e}}, R. 2005,
  \mnras, 363, 2

\bibitem[{{Kollmeier} {et~al.}(2014){Kollmeier}, {Weinberg}, {Oppenheimer},
  {Haardt}, {Katz}, {Dav{\'e}}, {Fardal}, {Madau}, {Danforth}, {Ford},
  {Peeples}, \& {McEwen}}]{kollmeier_underproduction}
{Kollmeier}, J.~A., {Weinberg}, D.~H., {Oppenheimer}, B.~D., {et~al.} 2014,
  \apjl, 789, L32

\bibitem[{{Lake} {et~al.}(2015){Lake}, {Zheng}, {Cen}, {Sadoun}, {Momose}, \&
  {Ouchi}}]{lake2015}
{Lake}, E., {Zheng}, Z., {Cen}, R., {et~al.} 2015, \apj, 806, 46

\bibitem[{{Liang} {et~al.}(2015){Liang}, {Kravtsov}, \& {Agertz}}]{Liang_2015}
{Liang}, C.~J., {Kravtsov}, A.~V., \& {Agertz}, O. 2015, ArXiv e-prints,
  arXiv:1507.07002

\bibitem[{{Marasco} {et~al.}(2015){Marasco}, {Debattista}, {Fraternali}, {van
  der Hulst}, {Wadsley}, {Quinn}, \& {Ro{\v s}kar}}]{Marasco_2015}
{Marasco}, A., {Debattista}, V.~P., {Fraternali}, F., {et~al.} 2015, \mnras,
  451, 4223

\bibitem[{{Martin} {et~al.}(2014){Martin}, {Chang}, {Matuszewski}, {Morrissey},
  {Rahman}, {Moore}, \& {Steidel}}]{Martin_2014}
{Martin}, D.~C., {Chang}, D., {Matuszewski}, M., {et~al.} 2014, \apj, 786, 106

\bibitem[{{Matsuda} {et~al.}(2011){Matsuda}, {Yamada}, {Hayashino}, {Yamauchi},
  {Nakamura}, {Morimoto}, {Ouchi}, {Ono}, {Kousai}, {Nakamura}, {Horie},
  {Fujii}, {Umemura}, \& {Mori}}]{Matsuda_2011}
{Matsuda}, Y., {Yamada}, T., {Hayashino}, T., {et~al.} 2011, \mnras, 410, L13

\bibitem[{{Milliard} {et~al.}(2010){Milliard}, {Martin}, {Schiminovich},
  {Evrard}, {Matuszewski}, {Rahman}, {Tuttle}, {McLean}, {Deharveng}, {Mirc},
  {Grange}, \& {Chave}}]{Milliard_2010}
{Milliard}, B., {Martin}, D.~C., {Schiminovich}, D., {et~al.} 2010, in Society
  of Photo-Optical Instrumentation Engineers (SPIE) Conference Series, Vol.
  7732, Society of Photo-Optical Instrumentation Engineers (SPIE) Conference
  Series, 5

\bibitem[{{Nelson} {et~al.}(2015{\natexlab{a}}){Nelson}, {Genel}, {Pillepich},
  {Vogelsberger}, {Springel}, \& {Hernquist}}]{Nelson_2015b}
{Nelson}, D., {Genel}, S., {Pillepich}, A., {et~al.} 2015{\natexlab{a}}, ArXiv
  e-prints, arXiv:1503.02665

\bibitem[{{Nelson} {et~al.}(2015{\natexlab{b}}){Nelson}, {Genel},
  {Vogelsberger}, {Springel}, {Sijacki}, {Torrey}, \&
  {Hernquist}}]{Nelson_2015a}
{Nelson}, D., {Genel}, S., {Vogelsberger}, M., {et~al.} 2015{\natexlab{b}},
  \mnras, 448, 59

\bibitem[{{Oppenheimer} \& {Dav{\'e}}(2008)}]{Oppenheimer_2008}
{Oppenheimer}, B.~D., \& {Dav{\'e}}, R. 2008, \mnras, 387, 577

\bibitem[{{Oppenheimer} {et~al.}(2016){Oppenheimer}, {Crain}, {Schaye},
  {Rahmati}, {Richings}, {Trayford}, {Tumlinson}, {Bower}, {Schaller}, \&
  {Theuns}}]{oppenheimer_2016}
{Oppenheimer}, B.~D., {Crain}, R.~A., {Schaye}, J., {et~al.} 2016, \mnras,
  arXiv:1603.05984

\bibitem[{{Prochaska} {et~al.}(2011){Prochaska}, {Weiner}, {Chen}, {Mulchaey},
  \& {Cooksey}}]{Prochaska_2011}
{Prochaska}, J.~X., {Weiner}, B., {Chen}, H.-W., {Mulchaey}, J., \& {Cooksey},
  K. 2011, \apj, 740, 91

\bibitem[{{Putman} {et~al.}(2012){Putman}, {Peek}, \& {Joung}}]{Putman_review}
{Putman}, M.~E., {Peek}, J.~E.~G., \& {Joung}, M.~R. 2012, \araa, 50, 491

\bibitem[{{Roy} {et~al.}(2015){Roy}, {Nath}, \& {Sharma}}]{roy_escape}
{Roy}, A., {Nath}, B.~B., \& {Sharma}, P. 2015, \mnras, 451, 1939

\bibitem[{{Salem} \& {Bryan}(2014)}]{Salem_2014a}
{Salem}, M., \& {Bryan}, G.~L. 2014, \mnras, 437, 3312

\bibitem[{{Salem} {et~al.}(2015){Salem}, {Bryan}, \& {Corlies}}]{Salem_2015}
{Salem}, M., {Bryan}, G.~L., \& {Corlies}, L. 2015, ArXiv e-prints,
  arXiv:1511.05144

\bibitem[{{Salem} {et~al.}(2014){Salem}, {Bryan}, \& {Hummels}}]{Salem_2014b}
{Salem}, M., {Bryan}, G.~L., \& {Hummels}, C. 2014, \apjl, 797, L18

\bibitem[{{Schaye} {et~al.}(2010){Schaye}, {Dalla Vecchia}, {Booth}, {Wiersma},
  {Theuns}, {Haas}, {Bertone}, {Duffy}, {McCarthy}, \& {van de
  Voort}}]{schaye_owlsim}
{Schaye}, J., {Dalla Vecchia}, C., {Booth}, C.~M., {et~al.} 2010, \mnras, 402,
  1536

\bibitem[{{Schaye} {et~al.}(2015){Schaye}, {Crain}, {Bower}, {Furlong},
  {Schaller}, {Theuns}, {Dalla Vecchia}, {Frenk}, {McCarthy}, {Helly},
  {Jenkins}, {Rosas-Guevara}, {White}, {Baes}, {Booth}, {Camps}, {Navarro},
  {Qu}, {Rahmati}, {Sawala}, {Thomas}, \& {Trayford}}]{schaye_eagle}
{Schaye}, J., {Crain}, R.~A., {Bower}, R.~G., {et~al.} 2015, \mnras, 446, 521

\bibitem[{{Shull} {et~al.}(2015){Shull}, {Moloney}, {Danforth}, \&
  {Tilton}}]{shull_2015}
{Shull}, J.~M., {Moloney}, J., {Danforth}, C.~W., \& {Tilton}, E.~M. 2015,
  \apj, 811, 3

\bibitem[{{Sijacki} {et~al.}(2007){Sijacki}, {Springel}, {Di Matteo}, \&
  {Hernquist}}]{Sijacki_2007}
{Sijacki}, D., {Springel}, V., {Di Matteo}, T., \& {Hernquist}, L. 2007,
  \mnras, 380, 877

\bibitem[{{Simcoe} {et~al.}(2004){Simcoe}, {Sargent}, \& {Rauch}}]{Simcoe_2004}
{Simcoe}, R.~A., {Sargent}, W.~L.~W., \& {Rauch}, M. 2004, \apj, 606, 92

\bibitem[{{Simpson} {et~al.}(2015){Simpson}, {Bryan}, {Hummels}, \&
  {Ostriker}}]{Simpson_2015}
{Simpson}, C.~M., {Bryan}, G.~L., {Hummels}, C., \& {Ostriker}, J.~P. 2015,
  \apj, 809, 69

\bibitem[{{Smith} {et~al.}(2008){Smith}, {Sigurdsson}, \& {Abel}}]{smith_2008}
{Smith}, B., {Sigurdsson}, S., \& {Abel}, T. 2008, \mnras, 385, 1443

\bibitem[{{Sravan} {et~al.}(2015){Sravan}, {Faucher-Giguere}, {van de Voort},
  {Keres}, {Muratov}, {Hopkins}, {Feldmann}, {Quataert}, \&
  {Murray}}]{Sravan_2015}
{Sravan}, N., {Faucher-Giguere}, C.-A., {van de Voort}, F., {et~al.} 2015,
  ArXiv e-prints, arXiv:1510.06410

\bibitem[{{Steidel} {et~al.}(2000){Steidel}, {Adelberger}, {Shapley},
  {Pettini}, {Dickinson}, \& {Giavalisco}}]{Steidel_2000}
{Steidel}, C.~C., {Adelberger}, K.~L., {Shapley}, A.~E., {et~al.} 2000, \apj,
  532, 170

\bibitem[{{Steidel} {et~al.}(2011){Steidel}, {Bogosavljevi{\'c}}, {Shapley},
  {Kollmeier}, {Reddy}, {Erb}, \& {Pettini}}]{Steidel_2011}
{Steidel}, C.~C., {Bogosavljevi{\'c}}, M., {Shapley}, A.~E., {et~al.} 2011,
  \apj, 736, 160

\bibitem[{{Steidel} {et~al.}(2010){Steidel}, {Erb}, {Shapley}, {Pettini},
  {Reddy}, {Bogosavljevi{\'c}}, {Rudie}, \& {Rakic}}]{Steidel_2010}
{Steidel}, C.~C., {Erb}, D.~K., {Shapley}, A.~E., {et~al.} 2010, \apj, 717, 289

\bibitem[{{Suresh} {et~al.}(2015{\natexlab{a}}){Suresh}, {Bird},
  {Vogelsberger}, {Genel}, {Torrey}, {Sijacki}, {Springel}, \&
  {Hernquist}}]{Suresh_2015}
{Suresh}, J., {Bird}, S., {Vogelsberger}, M., {et~al.} 2015{\natexlab{a}},
  \mnras, 448, 895

\bibitem[{{Suresh} {et~al.}(2015{\natexlab{b}}){Suresh}, {Rubin}, {Kannan},
  {Werk}, {Hernquist}, \& {Vogelsberger}}]{Suresh_2015b}
{Suresh}, J., {Rubin}, K.~H.~R., {Kannan}, R., {et~al.} 2015{\natexlab{b}},
  ArXiv e-prints, arXiv:1511.00687

\bibitem[{{Swinbank} {et~al.}(2015){Swinbank}, {Vernet}, {Smail}, {De Breuck},
  {Bacon}, {Contini}, {Richard}, {R{\"o}ttgering}, {Urrutia}, \&
  {Venemans}}]{Swinbank_2015}
{Swinbank}, A.~M., {Vernet}, J.~D.~R., {Smail}, I., {et~al.} 2015, \mnras, 449,
  1298

\bibitem[{{Thom} \& {Chen}(2008)}]{Thom_2008}
{Thom}, C., \& {Chen}, H.-W. 2008, \apj, 683, 22

\bibitem[{{Trujillo-Gomez} {et~al.}(2015){Trujillo-Gomez}, {Klypin},
  {Col{\'{\i}}n}, {Ceverino}, {Arraki}, \& {Primack}}]{Trujillo_Gomez_2015}
{Trujillo-Gomez}, S., {Klypin}, A., {Col{\'{\i}}n}, P., {et~al.} 2015, \mnras,
  446, 1140

\bibitem[{{Tumlinson} {et~al.}(2011){Tumlinson}, {Thom}, {Werk}, {Prochaska},
  {Tripp}, {Weinberg}, {Peeples}, {O'Meara}, {Oppenheimer}, {Meiring}, {Katz},
  {Dav{\'e}}, {Ford}, \& {Sembach}}]{tumlinson_OVI}
{Tumlinson}, J., {Thom}, C., {Werk}, J.~K., {et~al.} 2011, Science, 334, 948

\bibitem[{{Tumlinson} {et~al.}(2013){Tumlinson}, {Thom}, {Werk}, {Prochaska},
  {Tripp}, {Katz}, {Dav{\'e}}, {Oppenheimer}, {Meiring}, {Ford}, {O'Meara},
  {Peeples}, {Sembach}, \& {Weinberg}}]{Tumlinson_HI}
---. 2013, \apj, 777, 59

\bibitem[{{Turk} {et~al.}(2011){Turk}, {Smith}, {Oishi}, {Skory}, {Skillman},
  {Abel}, \& {Norman}}]{yt}
{Turk}, M.~J., {Smith}, B.~D., {Oishi}, J.~S., {et~al.} 2011, \apjs, 192, 9

\bibitem[{{Turner} {et~al.}(2015){Turner}, {Schaye}, {Steidel}, {Rudie}, \&
  {Strom}}]{Turner_2015}
{Turner}, M.~L., {Schaye}, J., {Steidel}, C.~C., {Rudie}, G.~C., \& {Strom},
  A.~L. 2015, \mnras, 450, 2067

\bibitem[{{van de Voort} \& {Schaye}(2013)}]{vandeVoort_2013}
{van de Voort}, F., \& {Schaye}, J. 2013, \mnras, 430, 2688

\bibitem[{{Werk} {et~al.}(2013){Werk}, {Prochaska}, {Thom}, {Tumlinson},
  {Tripp}, {O'Meara}, \& {Peeples}}]{werk13}
{Werk}, J.~K., {Prochaska}, J.~X., {Thom}, C., {et~al.} 2013, \apjs, 204, 17

\bibitem[{{Werk} {et~al.}(2014){Werk}, {Prochaska}, {Tumlinson}, {Peeples},
  {Tripp}, {Fox}, {Lehner}, {Thom}, {O'Meara}, {Ford}, {Bordoloi}, {Katz},
  {Tejos}, {Oppenheimer}, {Dav{\'e}}, \& {Weinberg}}]{werk14}
{Werk}, J.~K., {Prochaska}, J.~X., {Tumlinson}, J., {et~al.} 2014, \apj, 792, 8

\bibitem[{{Wiersma} {et~al.}(2010){Wiersma}, {Schaye}, {Dalla Vecchia},
  {Booth}, {Theuns}, \& {Aguirre}}]{Wiersma_2010}
{Wiersma}, R.~P.~C., {Schaye}, J., {Dalla Vecchia}, C., {et~al.} 2010, \mnras,
  409, 132

\bibitem[{{Wise} \& {Cen}(2009)}]{wise_escape2}
{Wise}, J.~H., \& {Cen}, R. 2009, \apj, 693, 984

\bibitem[{{Wise} {et~al.}(2014){Wise}, {Demchenko}, {Halicek}, {Norman},
  {Turk}, {Abel}, \& {Smith}}]{wise_escape}
{Wise}, J.~H., {Demchenko}, V.~G., {Halicek}, M.~T., {et~al.} 2014, \mnras,
  442, 2560

\bibitem[{{Zheng} {et~al.}(2011){Zheng}, {Cen}, {Weinberg}, {Trac}, \&
  {Miralda-Escud{\'e}}}]{Zheng_2011}
{Zheng}, Z., {Cen}, R., {Weinberg}, D., {Trac}, H., \& {Miralda-Escud{\'e}}, J.
  2011, \apj, 739, 62

\end{thebibliography}

\appendix

\section{Uncertainty in the Extragalactic Ultraviolet Background}
The extragalactic ultraviolet background (EUVB) is an important component of any photoionization model implemented for both observations and simulations. However, the EUVB is historically not well constrained.  Variations of the \citet{HM96} background (e.g. HM96, HM01, HM05, HM12) are implemented in the majority of hydrodynamical simulations and in \textsc{cloudy}. Yet simulations attempting to match measurement of the low-redshift Lyman $\alpha$ forest find that the most recent HM12 model does not reproduce the observed column density distribution of absorbers. The simulations of \citet{kollmeier_underproduction} require a photoionization rate a factor of five higher than the HM12 model while those of \citet{shull_2015} suggest a factor of 2-3 increase.  However, unlike the previous work, \citet{shull_2015} find that the HM12 model does in fact reproduce the distribution of absorbers with log(N$_{\mathrm{H{\scriptsize I}}}$) $> 14.0$ and that no single model reproduces the entire distribution.  Both of these values are more consistent with HM96 and in-line with our high EUVB model, g1q10.  Finally, at lower redshifts, constraints are even harder to place on the EUVB \citep{cooray_2016}.  In contrast to both of these high redshift studies, \citet{adams_2011} placed upper limits on the photoionization rate at $z=0$ from the non-detection of H$\alpha$ in UGC 1281 at roughly the values in the HM12 model and in UGC 7321 at roughly 10 times lower  - corresponding to 10 and 100 times lower than the HM05 model assumed throughout this paper. This is in the range of our low model, g1q01. In this way, the models used here bracket the range of possible photoionization rates as they are known today.  These theoretical and observational inconsistencies highlight the uncertainty in the shape and intensity of the EUVB.

In addition, these models depend on the escape fraction of ionizing photons from their host galaxy, a quantity that is expected to be low but has been argued to be anywhere from 0.01 to 0.3, depending on the galaxy mass and redshift \citep{dove_escape,wise_escape2,benson_escape,roy_escape}. \citep[see introduction of ][for an in-depth discussion.]{wise_escape}  Both of the above Lyman $\alpha$ studies agree that the likely source of the discrepancy is the prescription for the escape fraction which leads to a galactic contribution to the EUVB that is too low at low-redshift. 

\begin{figure}[h]
\centering
\includegraphics[width=0.475\textwidth]{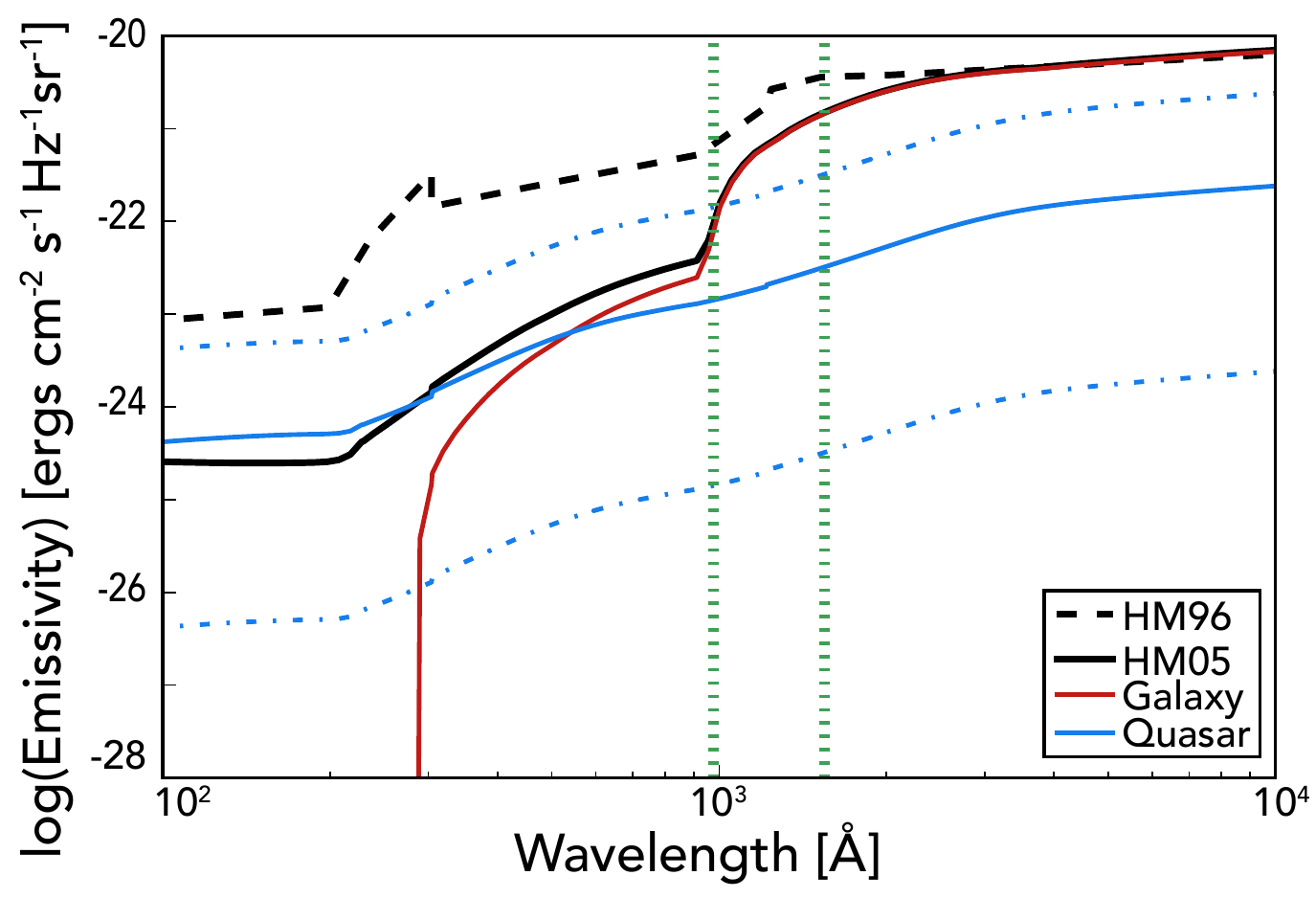}
\caption{Relevant EUVB backgrounds for this work: HM96 (assumed in the simulation) and HM05 (fiducial for \textsc{cloudy} modeling). The galaxy (red) and quasar (blue) components of HM05 are also plotted. The blue dashed lines represent the two quasar backgrounds assumed throughout the paper (100 times less intense and 10 times more intense respectively).  The green vertical lines bracket the wavelengths of the emission examined in the paper. \label{HM_background.fig}}
\end{figure}

Figure \ref{HM_background.fig} shows the EUVB relevant for this work. HM96 is the background assumed within the simulation and used for the chemical network that determined the cooling rate at each time step. HM05 is the background assumed and varied for all of the \textsc{cloudy} modeling discussed in the paper. Figure \ref{HM_background.fig} also shows how the HM05 background is broken into its components: the galactic (red) and the quasar (blue). Throughout the paper, this background is modified to either g1q01 (a quasar intensity 100 times smaller than fiducial) or g1q10 (a quasar intensity 10 times larger than fiducial).  These models are plotted as blue dashed-dotted lines. 

One concern is that the model assumed for the simulation (HM96) is not the same as the one assumed in the \textsc{cloudy} modeling done in this paper (HM05). At longer wavelengths, the HM96 and HM05 backgrounds are similar; however, the HM96 background is closer to the modified g1q10 background than to the fiducial HM05 at shorter wavelengths where the quasar component dominates. This suggests that the range of backgrounds being explored is reasonable. Furthermore, the only way the simulation directly depends on the EUVB is in the calculation of the heating and cooling. The heating will be dominated by physical processes such as supernova feedback.  As for the cooling, for $T> 10^4$K, which is the case for all CGM gas considered, \textsc{cloudy} modeling shows that the cooling function assumed in the simulation varies somewhat with the ionization fraction at the low metallicities found in the CGM but is dominated by the overall metallicity.  Thus pairing the simulated density and temperature with the varying EUVB in the \textsc{cloudy} modeling is not unreasonable.

The green vertical lines bracket the wavelength range of the emission lines considered here. They fall within the galaxy-dominated part of the spectrum. However, the quasar component contributes much more of the ionizing intensity and thus is more important in shaping the expected ion fractions for the column density and emission predictions. \citet{werk14} examine how differences between HM01 and HM12 affect their measurements and find that repeating the analysis with HM12 lowers the gas ionization parameters by 0.1-0.4 dex, which must be accounted for in either the H{\scriptsize I} column density or the metallicity. The simulated column densities show larger variations with changes in EUVB because the density and temperature are fixed and only the ionizing intensity is changing. In the less constrained \textsc{cloudy} modeling, flexibility within setting the interdependent quantities of ionization parameter and metallicity can reduce the effect of the EUVB. The value of the simulation is that these gas properties are determined by the larger cosmological context instead of modeling an isolated cloud.

Finally, in addition to the uncertainty in the EUVB, ionizing photons from local stellar sources are expected to be the dominant source of photoionization in local star forming regions within the disk but again, the escape fraction of these photons into the halo is entirely uncertain.  Similarly, star formation in the halo has been shown to change the extent and shape of a galaxy's Lyman $\alpha$ emission but this triggered emission results in a greater predicted UV flux than what is currently measured. \citep{lake2015}.  Because of these uncertainties and because we are focused on emission from gas further from the star-forming disk, including this ionization source is reserved for future work.

\section{Investigating the Effects of Resolution}
Because the column density and surface brightness calculations depend so sensitively on the density, temperature, and metallicity of the gas, understanding how the resolution of the simulation affects these quantities is necessary to evaluate their robustness.  On one hand, over-cooling can lead to large, artificially dense clumps and is known to leave a too centrally concentrated disk. On the other hand, observational evidence suggests that absorbing clouds can be small, high-density structures that would be under-resolved in the simulation \citep{crighton}.

\begin{figure}[h]
\centering
\includegraphics[width=0.55\textwidth]{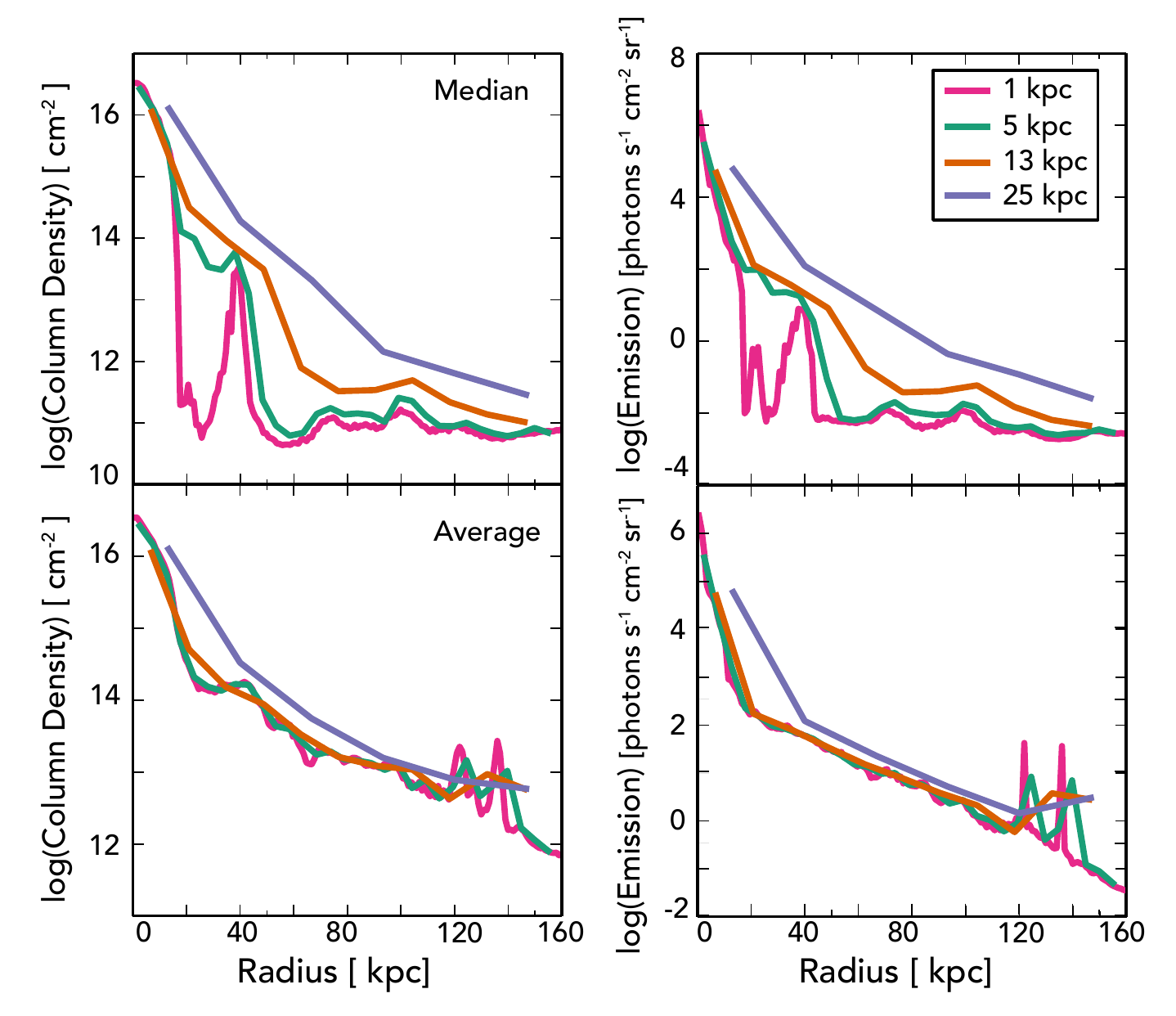}
\caption{Median and average column density and emission profiles of C{\scriptsize III} at $z=0.2$, binned for four different resolutions. 1 kpc is the resolution assumed throughout the paper and is roughly the underlying simulation resolution beyond the disk in the CGM.  As the resolution increases, the median profile decreases as the gas structure is refined. At the very center of the disk and in the outer halo, the median profile of the simulation appears to be converging below 5 kpc.  The exception is the disk-halo interface at roughly 20 kpc. Only at the highest resolution is the sharp transition from disk to halo captured. The average converges more quickly.} \label{resolution.fig}
\end{figure}

The ideal solution would be to re-run the simulation at lower resolution and compare its output to the simulation analyzed here.  Unfortunately, because of the length of time that has passed between this analysis and the original, high-resolution run, the exact initial conditions of this simulation can not be reproduced. However, we can attempt to address this issue by re-binning the high resolution output to lower resolutions. Figure \ref{resolution.fig} shows the median and average profiles of the column density and emission of C{\scriptsize III} at $z=0.2$ for the volume considered in the above analysis. The average profiles of these quantities have converged except at the lowest resolution.  On the other hand, at lower resolutions, the median profiles of the column density and the emission are both too high, corresponding to higher average density and fewer low-density regions.  However, by a resolution of 5 kpc, the simulation seems to have converged on a median profile for the inner disk as well as the outer parts of the halo. The remaining region, around 20 kpc, corresponds to the edge of the disk-halo interface. It's only at the highest resolution that this interface is properly resolved. At 5 kpc, this boundary is still blurred, not allowing for the sharp transition from high-density, cold gas to low-density, warm gas.

These profiles suggest that further resolving the outer halo should not greatly change the median predictions for the column density and emission.  While higher densities clumps are mostly likely still not being resolved even with the maximum resolution, Figure \ref{resolution.fig} suggests that these regions are small compared to the volume and won't significantly alter the median profiles of either quantity.

\end{document}